\documentclass[twocolumn,pre,aps,superscriptaddress,reprint]{revtex4} 
\usepackage{subfigure}
\usepackage{psfrag,graphicx}
\usepackage{pict2e}
\usepackage{dcolumn}
\usepackage{bm}
\usepackage{latexsym}
\usepackage{epstopdf}
\usepackage{enumerate}
\usepackage[english]{babel}
\usepackage{mathtools}

\usepackage{bbold}
\usepackage{amssymb, bigints, amsmath, braket}   
\usepackage[usenames, dvipsnames]{color}
\usepackage{float}
\DeclareMathOperator{\Tr}{Tr}

\usepackage{xcolor}
\usepackage{simplewick}
\usepackage[normalem]{ulem}
\newcommand{\stkout}[1]{\ifmmode\text{\sout{\ensuremath{#1}}}\else\sout{#1}\fi}

\definecolor{myred}{rgb}{1,0.,0.3}
\definecolor{myblue}{named}{MidnightBlue}
\usepackage[colorlinks=true,citecolor=myblue,linkcolor=myred]{hyperref}

\begin{document}

\title{Quantum chaotic fluctuation-dissipation theorem: Effective Brownian motion in closed quantum systems}
\author{Charlie Nation}%
 \email{C.Nation@sussex.ac.uk}
\affiliation{%
Department of Physics and Astronomy, University of Sussex, Brighton, BN1 9QH, United Kingdom.
}%
\author{Diego Porras}%
 \email{D.Porras@iff.csic.es}
 \affiliation{%
 Department of Physics and Astronomy, University of Sussex, Brighton, BN1 9QH, United Kingdom.
 }%
\affiliation{%
 Institute of Fundamental Physics IFF-CSIC, Calle Serrano 113b, 28006 Madrid, Spain 
} 
\date{\today}

\begin{abstract}
We analytically describe the decay to equilibrium of generic observables of a non-integrable system after a perturbation in the form of a random matrix. We further obtain an analytic form for the time-averaged fluctuations of an observable in terms of the rate of decay to equilibrium. Our result shows the emergence of a Fluctuation-Dissipation theorem corresponding to a classical Brownian process, specifically, the Ornstein-Uhlenbeck process. Our predictions can be tested in quantum simulation experiments, thus helping to bridge the gap between theoretical and experimental research in quantum thermalization. We test our analytic results by exact numerical experiments in a spin-chain. We argue that our Fluctuation-Dissipation relation can be used to measure the density of states involved in the non-equilibrium dynamics of an isolated quantum system.
\end{abstract}

\maketitle

\section{Introduction} %
Ubiquitous to nearly all fields of the natural sciences is the phenomenon of equilibration to a thermal state. 
However, in the context of quantum systems a full understanding of thermalization has remained enigmatic. 
This long-studied problem \cite{Neumann2010} has seen a resurgence of interest in recent years \cite{Rigol2008,DAlessio2016,Gogolin2016,Borgonovi2016,Mori2018}, 
largely driven by the modern experimental capability to study the unitary quantum dynamics of closed systems \cite{Schreiber2015,Clos2016a,Kaufman2016, Neill2016, Neill2018}. Of particular interest is the thermalization of initial pure-states, which cannot easily be expected to equilibrate to some statistical ensemble. This is the case treated in the present work.

On the theoretical side there have been advances in two key areas: Typicality, and the Eigenstate Thermalization Hypothesis (ETH). The typicality approach has shown that most pure states of a large system correspond to a local canonical ensemble in some small (with respect to the total system size) subspace\cite{Popescu2006, Reimann2008, Linden2009, Bartsch2009}, whilst the ETH has provided a mechanism for thermalization - the eigenstates themselves form an effective microcanonical ensemble \cite{Deutsch1991,Srednicki1999}. This has been supported by a large amount of numerical evidence \cite{Santos2010,Santos2010a,Steinigeweg2013,Beugeling2014,Beugeling2015,Hunter-Jones2018, Mondaini2016, Yoshizawa2018}.

Despite much recent progress on the understanding of thermalization, there has been less work describing the decay process \cite{Reimann2016, Borgonovi2018, Richter2018} or the timescales of equilibration \cite{Garcia-Pintos2015}. 
We address both of these using a Random Matrix Theory (RMT) model \cite{Deutsch1991, Reimann2015}, which the current authors have recently shown reproduces the ETH ansatz \cite{Nation2018}. We describe the decay to equilibrium of generic non-integrable quantum systems, and obtain an expression for the time-averaged fluctuations of local observables in terms of their rate of decay to equilibrium; thus observing an emergent classical Fluctuation-Dissipation Theorem (FDT), analogous to those derived from a Langevin equation for Brownian motion. 

FDTs describe a relationship between the linear response of a system to some perturbation and its fluctuations in thermal equilibrium \cite{Kubo1966}.  An example that is particularly relevant for this work is the case of an Ornstein-Uhlenbeck process. This is a Brownian process with diffusion constant $D$, where particle positions are additionally subjected to a deterministic drift of the form 
$\dot{x} = -\gamma x$. 
The particle position is a stochastic variable whose time-averaged fluctuations satisfy the relation \cite{Breuer2002},
\begin{equation}\label{eq:Einstein}
  \langle x^2 \rangle = \frac{D}{\gamma}.
\end{equation}
In this work we show that the fluctuations of the expectation value of a local operator, $\langle O(t) \rangle$, of a quantum chaotic system follows a similar relation, with $D$ replaced by the inverse of the density of states (DOS).  Our result radically differs from previous theoretical results linking the {\it quantum} FDT for quantum fluctuations $\langle \Delta O^2 \rangle$ \cite{Khatami2013} to linear response theory and the ETH \cite{Srednicki1999}.

This article is arranged as follows. In section II we outline the physical scenario in question, our RMT approach \cite{Nation2018}, our key assumptions and their justifications, and how the the ETH may be derived, and exploited, from our methods. In section III we derive our main analytical result - an explicit expression for the equilibration in time of generic observables towards their microcanonical average. In section IV, we see that exploiting a result from \cite{Nation2018}, the results of section III provide a FDT for chaotic quantum systems. To confirm the applicability of our RMT description to realistic physical models, in section V we present exact diagonalization calculations of a quantum spin-chain, and apply this to a generalized FDT section VI. In section VII we propose and numerically simulate an approach to experimental verification of our findings. Finally, we conclude in Section VIII. Various details and derivations are provided in Appendices.

\section{Random Matrix Theory approach to quantum thermalization}

\subsection{Physical Scenario} 
Our objective is to analyze the quantum dynamics of a many-body system whose total Hilbert space, ${\cal H}$, is divided into two subspaces, 
${\cal H} = {\cal H}_S \otimes {\cal H}_B$. 
${\cal H}_S$ is a local Hilbert space corresponding, for example, to one or a few sites in a quantum lattice system. 
${\cal H}_B$ is a larger Hilbert space which will play the role of a finite many-body quantum bath. 

\begin{figure}
	\includegraphics[width=\linewidth]{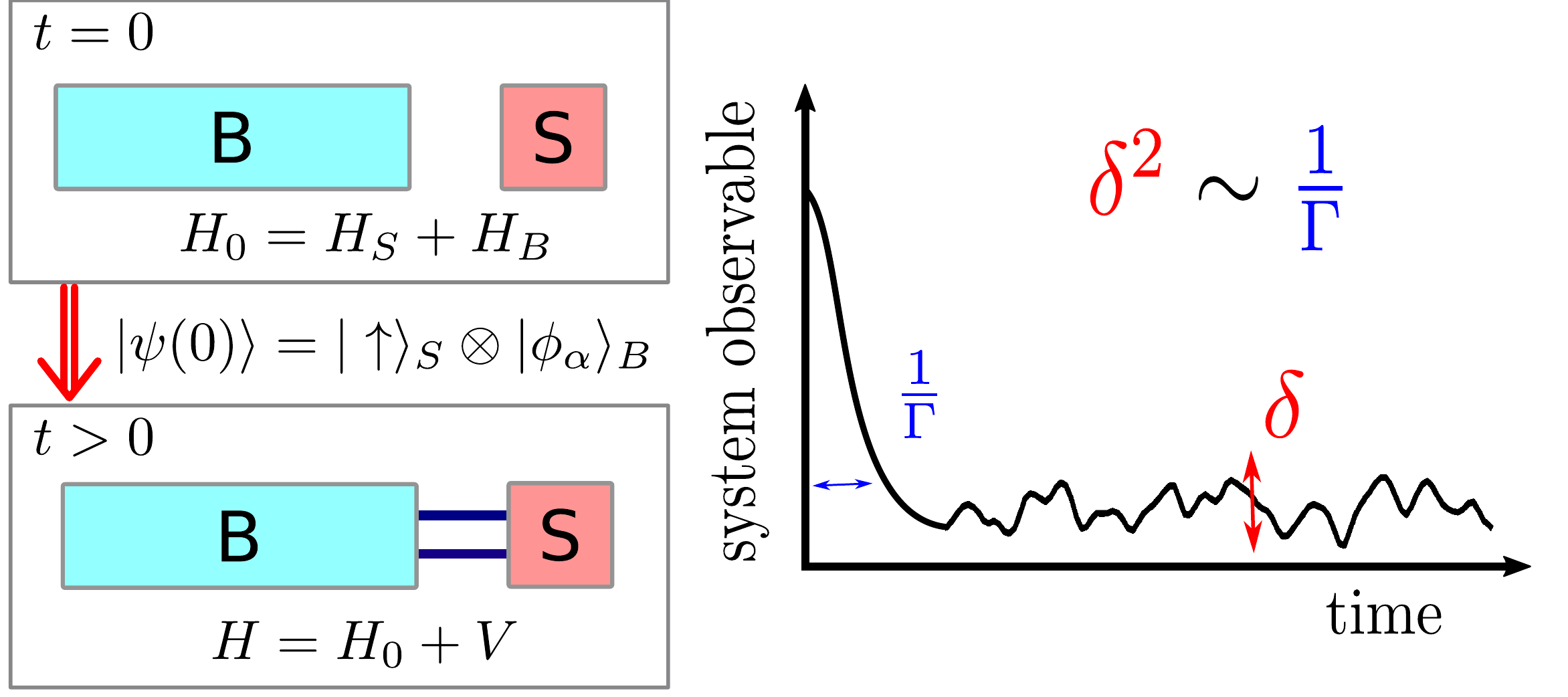}
	\caption{Diagram depicting quench at $t = 0$ from $H_0 = H_S + H_B$ to $H = H_0 + V$, where $V$ couples the system and bath. The initial state is an eigenstate $|\phi_\alpha\rangle$ of $H_{ \rm 0}$ (this condition is relaxed below).}
	\label{fig:diagram}
\end{figure}

We investigate the case in which a non-interacting Hamiltonian of the form $H_0 = H_S + H_B$, is perturbed by a term $V$ to form a fully interacting Hamiltonian,
\begin{equation}
H = H_0 + V.
\label{quench}
\end{equation}
$H_S$ and $H_B$ in $H_0$ act on Hilbert subspaces ${\cal H}_S$ and ${\cal H}_B$, respectively, and 
$V$ is an interaction term between the system and the bath. The simplest situation that we will consider is a quantum quench scenario, in which the system is initially in an eigenstate of $H_0$ at $t = 0$, as illustrated in Fig. (\ref{fig:diagram}). 
We will see, however, that this assumption on the initial state can be relaxed under certain conditions. 
The goal of this work is to understand the general properties of the dynamics of an observable $O$ acting on 
${\cal H}_S$.

In a non-integrable system a qualitative description is obtained by replacing the coupling $V$ by a random matrix. 
Typically, $V$ is the sum of a few products of local operators which takes the form  
$V = \sum_n g_n O_{S,n}  O_{B,n}$, where $O_{S,n}$ are local operators acting on ${\cal H}_S$, and $O_{B,n}$ are local operators acting on ${\cal H}_B$. 
If the bath Hamiltonian, $H_B$, is non-integrable, we expect that operators $O_{B,n}$ are well described by Gaussian random matrices (see for example \cite{Beugeling2015,Mondaini2016} for a recent numerical confirmation), and as such a random matrix ansatz should also be a good approximation for $V$. 

Throughout this work we will consider a weak coupling limit, such that we can assume that the 
random matrix $V$ is homogeneous. 
In general, one may expect that the coupling matrix $V$ has some structure, for example, matrix elements $V_{\alpha\beta}$ typically decay as a function of the energy difference between states $\alpha$ and $\beta$. A reasonable assumption is to consider that the matrix elements of $V$ are constant within a typical energy band of width $\Gamma_{V}$. The approximation of $V$ as a homogeneous Gaussian random matrix will be justified as long as 
$\Gamma \ll \Gamma_{V}$, where $\Gamma$ is the energy scale associated to the system-bath coupling. 
The weak coupling limit can be satisfied in the case that $H_S$ describes an impurity weakly coupled to a many-body bath described by $H_B$. This limit is, however, not trivially fulfilled in the case that $H$ represents a homogeneous system. 
In this case, $\Gamma_V$ and $\Gamma$ could be of similar magnitude, since 
$\Gamma_V$ is associated to interactions in $H_B$, which in a homogeneous system would be similar in magnitude to the coupling term 
$V$. As explained in Appendix \ref{App:Summary}, our theory and general results could be modified to account for this situation.

\subsection{Random Matrix Model} 
The random matrix model under study is that used in the pioneering work of Deutsch \cite{Deutsch1991}. The spirit of this approach is to model both $H_0$ and $V$, as well as operators describing local observables, by matrices that have the same properties as the equivalent operators in physical systems.

The non-interacting part in \eqref{quench}, $H_0$, is modelled by a diagonal matrix of size $N$, with $N$ the total dimension of the Hilbert space,
\begin{equation}
(H_0)_{\alpha \beta} = E_\alpha \delta_{\alpha \beta}
\end{equation}
where $E_\alpha = \alpha \omega_0$, and $\omega_0 = 1/N$ is the spacing between energy levels, which is assumed to be constant. This approximation will be relaxed later on by assuming an energy-dependent density of states. The perturbation term in Eq. \eqref{quench} is modelled by a random matrix,
\begin{equation}\label{eq:Hamiltonian}
V_{\alpha \beta}  = h_{\alpha\beta},
\end{equation}
where $h_{\alpha\beta}$ are independent random numbers selected from the Gaussian Orthogonal Ensemble 
(GOE), such that the matrix $h$ has the probability distribution, 
\begin{equation}
P(h) \propto \exp\left[-\frac{N}{4 g^2} \Tr h^2\right],
\end{equation}
giving $\langle h_{\alpha\beta} \rangle = 0$, and $\langle h_{\alpha\beta}^2 \rangle = g^2 / N$ for $\alpha \neq \beta$, and otherwise $\langle h_{\alpha\alpha}^2 \rangle = 2g^2 / N$. 

From here on we denote the set of eigenstates of $H$ (interacting basis) by $\{|\psi_\mu\rangle\}$,
\begin{equation}
H |\psi_\mu \rangle = E_\mu |\psi_\mu\rangle, \ \ \mu = 1, 2, \dots, N,
\end{equation}
and the eigenstates of $H_0$  (non-interacting basis) by $\{|\phi_\alpha\rangle\}$
\begin{equation}
H_0 |\phi_\alpha \rangle = E_\alpha |\phi_\alpha\rangle, \ \ \alpha = 1, 2, \dots, N.
\end{equation}
We can approximate $E_\mu = \mu \omega_0$, since the perturbation is homogeneous, and thus will not change the average spacing between energy levels. 
To simplify the notation, we always refer to the non-interacting basis (interacting basis) by indexes with Greek letters $\alpha$, $\beta$, ($\mu$, $\nu$). 
Sums over wavefunction indices in expressions below are always understood to run over values $1, 2, \dots, N$.

We define the interacting wavefunctions, $c_\mu(\alpha)$, 
\begin{equation}
|\psi_\mu\rangle = \sum_{\alpha}c_\mu(\alpha)|\phi_\alpha\rangle,
\end{equation}
where $c_\mu(\alpha)$ are random variables whose statistical properties depend on the properties of the random matrix $V$.
Deutsch \cite{Deutscha} obtained an expression for the probability distribution of eigenstates,
\begin{equation}\label{eq:Lambda}
\langle|c_\mu(\alpha)|^2\rangle_V := \Lambda(\mu, \alpha) = \frac{\omega_0 \Gamma / \pi }{(E_\mu - E_\alpha)^2 + \Gamma^2},
\end{equation}
where $\Gamma = \frac{\pi g^2}{N \omega_0}$\cite{Footnote}, and $\langle \cdots \rangle_V$ denotes an average over realizations of the random perturbation $V$.  We assume a feature of large random matrices known as self-averaging, and replace summations over coefficients by their ensemble average, 
\begin{equation}\label{eq:self_averaging}
\sum_{\alpha\cdots\beta}c_\mu(\alpha)\cdots c_\nu(\beta) \to \sum_{\alpha\cdots\beta}\langle c_\mu(\alpha)\cdots c_\nu(\beta) \rangle_V.
\end{equation}
This is a very common assumption in the treatment of random matrices \cite{Guhr1998}, and is well justified numerically for this model in \cite{Nation2018}. 

\subsection{Correlation functions of quantum chaotic wavefunctions}
The RMT approach will allow us to express the dynamics of local observables in a non-integrable system in terms of averages of products of random wavefunctions, $c_\mu(\alpha)$. 
At first sight, a reasonable approximation would be to consider that $c_\mu(\alpha)$ are independent Gaussian variables, such that any multi-point correlation function can be simply obtained as a product of two-point correlations for the form given by Eq. \eqref{eq:Lambda}. 

In Ref. \cite{Nation2018} the current authors elaborated further on a theoretical approach developed by J. Deutsch \cite{Deutsch1991}, and extended it to include the effect of the orthonormality between wavefunctions, which can be understood as an effective repulsive interaction in a statistical theory of the variables $c_\mu(\alpha)$. We showed that the inclusion of correlations between $c_\mu(\alpha)$ is essential to obtain the correct form of the ETH conjectured for off-diagonal elements of generic observables, in agreement with Srednicki's ansatz \cite{Srednicki1999}. 
We review this proof in detail in Appendices \ref{App:Summary} and \ref{App:ETH}, and discuss here the most relevant results. 

We focus here on two sets of correlation functions of interest: $\langle c_\mu(\alpha) c_\nu(\beta) c_\mu(\alpha^\prime) c_\nu(\beta^\prime) \rangle_V$ for both $\mu = \nu$, and $\mu \neq \nu$:
\begin{enumerate}[(i)]
\item For $\mu = \nu$ we can show that the orthonormality constraint does not affect the calculation, such that the coefficients may be treated as independent Gaussian variables,
\begin{equation}\label{eq:4pt_diag}
\begin{split}
\langle c_\mu(\alpha)& c_\mu(\beta) c_\mu(\alpha^\prime) c_\mu(\beta^\prime) \rangle_V = \\ &\Lambda(\mu, \alpha)\Lambda(\nu, \beta)\delta_{\alpha\alpha^\prime}\delta_{\beta\beta^\prime} \\
+&\Lambda(\mu, \alpha)\Lambda(\nu, \alpha^\prime)(\delta_{\alpha\beta}\delta_{\alpha^\prime\beta^\prime} +\delta_{\alpha\beta^\prime}\delta_{\beta\alpha^\prime} ).
\end{split}
\end{equation}
We will see that this term plays a role in the prediction of time-averages of expectation values of observables $\langle O(t) \rangle$.  We note that for this to reproduce the expected microcanonical average, the contributions of the latter two terms in Eq. \eqref{eq:4pt_diag} must be small, which is shown in Appendix \ref{App:ETH}.

\item For  $\mu \neq \nu$  we find
\begin{equation}\label{eq:4pt_off_diag}
\begin{split}
\langle & c_\mu(\alpha) c_\nu(\beta) c_\mu(\alpha^\prime) c_\nu(\beta^\prime) \rangle_V = \Lambda(\mu, \alpha)\Lambda(\nu, \beta)\delta_{\alpha\alpha^\prime}\delta_{\beta\beta^\prime} \\&
- \frac{\Lambda(\mu, \alpha)\Lambda(\nu, \beta)\Lambda(\mu, \alpha^\prime)\Lambda(\nu, \beta^\prime)}{\Lambda^{(2)}(\mu, \nu)}(\delta_{\alpha\beta}\delta_{\alpha^\prime\beta^\prime} +\delta_{\alpha\beta^\prime}\delta_{\beta\alpha^\prime} ).
\end{split}
\end{equation}
This case is especially relevant for non-equilibrium dynamics, as it dictates both the equilibrium fluctuations, as well as the decay to equilibrium of a given observable $O$. 
\end{enumerate}

\subsection{Assumptions on physical observables}\label{Sec:Assumptions}

A very non-trivial aspect of our theory is the introduction of matrices that model local observables in physical non-integrable systems. We impose two conditions on a Hermitian matrix, $O$, that are satisfied by local observables:

{\it Sparsity.-} We assume that, $O$, expressed in the non-interacting basis, is represented by a diagonal matrix in the non-interacting basis or, at least, by a matrix with only a few non-diagonal entries. 
This implies that matrix elements in the non-interacting basis, $O_{\alpha\beta} := \langle \phi_\alpha|O|\phi_\beta\rangle$, can be written like
\begin{equation}
O_{\alpha \beta} = \sum_{n \in {\cal N}_O} O_{\alpha,\alpha+n} \delta_{\beta, \alpha+n}
\end{equation}
where ${\cal N}_O$ is a set of $N_O$ integer values which determines the non-diagonal finite matrix elements. 
The sparsity constraint is satisfied if $N_O \ll N$.

In a physical system the sparsity condition is fulfilled as long as the observable $O$ is defined on the local Hilbert space ${\cal H}_S$. 
To see this more clearly, let us express the non-interacting basis in the form of products of eigenstates of $H_{S}$ and $H_{B}$. We define $|s \rangle_S$, with 
$s = 1, \dots, {\rm dim}({\cal H}_S)$, as the eigenstates of $H_S$ 
with energy $E_s^S$, 
and $|\alpha_B \rangle_B$, with $\alpha_B = 1, \dots, {\rm dim}({\cal H}_B)$, the set of eigenstates of $H_B$, with energy $E^B_{\alpha_B}$. 
An eigenstate of the non-interacting Hamiltonian is given by
\begin{equation}
| \phi_\alpha \rangle = | s(\alpha) \rangle_S | \alpha_B(\alpha) \rangle_B ,
\label{product.basis}
\end{equation}
where $s(\alpha)$ and $\alpha_B(\alpha)$ are the system and bath eigenstate number of the non-interacting state $\alpha$, respectively. The energy of $| \phi_\alpha \rangle$ is
\begin{equation}
E_\alpha = E^S_{s(\alpha)} + E^B_{\alpha_B(\alpha)}.
\end{equation}
A local operator will only couple states with different local quantum number $s$, and thus, $O_{\alpha \beta} \neq 0$ only if
\begin{equation}
E_\alpha - E_\beta = E^S_{s(\alpha)} - E^S_{s(\beta)}.
\end{equation}
In this case, $O$ induces transitions between only a few states that are separated by one of the possible gaps of $H_S$.
Consider for example that $H_S$ is a local term in a spin chain. 
Then a local operator, $O = \sigma_z$ or $\sigma_+$, $\sigma_-$ will induce transitions only between non-interacting states with a flipped local spin, such that $N_O = 3$.

{\it Smoothness.-} 
In the following calculations we will have to evaluate sums of observable matrix elements in the non-interacting basis weighted by probability distributions. For this we will define a smoothed version of the observable in the following way,
\begin{equation}
\overline{[O_{\alpha,\alpha+n}]}_{\mu} 
:= \sum_{\alpha}\Lambda(\mu, \, \alpha)O_{\alpha,\alpha+n}.
\label{def.average}
\end{equation}
The quantity $\overline{[O_{\alpha,\alpha+n}]}_{\mu}$ represents the average of non-interacting matrix elements along the $n$'th diagonal, weighted by the Lorentzian function \eqref{eq:Lambda}. 
We will refer to the quantity 
$\overline{[O_{\alpha,\alpha+n}]}_{\mu}$ as the 
microcanonical average of the matrix elements
$O_{\alpha,\alpha+n}$ around the energy $E_\mu$. 
This average is well defined as long as (see below),
\begin{eqnarray}
&& \frac{\Gamma}{\omega_0} \gg 1, \nonumber \\
&& \Gamma^2 
\frac{d^2}{d E_\mu^2} \overline{[O_{\alpha,\alpha+n}]}_{\mu} \ll 1.
\label{smoothness}
\end{eqnarray}
The first condition implies that a large number of matrix elements are averaged in the sum in Eq. \eqref{def.average}. 
The second conditions implies that the average $\overline{[O_{\alpha,\alpha+n}]}_{\mu}$ varies smoothly as function of the energy $E_\mu$. 

The smoothness conditions \eqref{smoothness} imply that, to a good approximation, we can substitute the matrix elements $O_{\alpha,\alpha+n}$ by their smoothed version,
$\overline{[O_{\alpha,\alpha+n}]}_{\mu}$, whenever matrix elements appear within summations over a large number of states. Imagine for example that we have a function 
$F_{\alpha_0}(\alpha)$, which is centred around $\alpha = \alpha_0$ and has an energy width $\Gamma_F$, when expressed as a function of $E_\alpha$.
The smoothness condition implies that
\begin{equation}
\sum_\alpha O_{\alpha \alpha} F_{\alpha_0}(\alpha) 
\approx  
\overline{[O_{\alpha \alpha}]}_{\alpha_0} \sum_\alpha 
F_{\alpha_0}(\alpha) ,
\label{smoothness.assumption}
\end{equation}
provided that the variation of  $\overline{[O_{\alpha \alpha}]}_{\alpha_0}$ as a function of $E_{\alpha_0}$ can be neglected within an energy interval of width $\Gamma_F$. 

In practice, in the following calculations, matrix elements will always be evaluated in products with functions of typical width $\Gamma$. Hence, we observe that averages such as Eq. \eqref{def.average} can be seen as a microcanonical averaging of the matrix elements $O_{\alpha,\alpha+n}$ around the central energy $E_{\overline{\mu}}$. Our calculations going forward require that this average changes slowly over the width $\Gamma$ of $\Lambda(\mu, \alpha)$.


Indeed, the conditions, \eqref{smoothness}, under which the smoothness assumption holds can be understood by considering the values $O_{\alpha\alpha}$ as random numbers with a mean value ${\cal O}(\alpha) = \overline{O_{\alpha\alpha}}$. This is obviously a rough approach
to the study of the values of an observable in the non-interacting
basis. However, this method will allow us to
understand the conditions under which the smoothness
assumption is satisfied.

Consider a certain probability function $p_{\alpha_0}(\alpha)$ centred around the value $\alpha_0$, normalized with a width $\gamma_p$ such that 
\begin{eqnarray}
\sum_\alpha p_{\alpha_0}(\alpha) &=& 1, \nonumber \\
\sum_\alpha p_{\alpha_0}(\alpha)(\alpha -\alpha_0) &=& 0, \nonumber \\
\sum_\alpha p_{\alpha_0}(\alpha)(\alpha -\alpha_0)^2 &=& \gamma_p .
\end{eqnarray}
We want to quantify to what extent the following approximation holds,
\begin{equation}\label{eq:smoothness_p_alpha}
\sum_{\alpha} O_{\alpha \alpha} p_{\alpha_0}(\alpha) \approx {\cal O}(\alpha_0).
\end{equation}
We thus calculate the variance
\begin{equation}\label{eq:var_micro}
\Delta^2_O = 
\overline{
	\left( 
	\sum_{\alpha} O_{\alpha \alpha} p_{\alpha_0}(\alpha) - {\cal O}(\alpha_0) \right)^2 }.
\end{equation}
Now, by expanding Eq. \eqref{eq:var_micro}, 
\begin{equation}
\begin{split}
\Delta_O^2 & = \sum_{\substack{\alpha\beta \\ \alpha\neq\beta}} \overline{O_{\alpha\alpha}O_{\beta\beta}}p_{\alpha_0}(\alpha)p_{\alpha_0}(\beta) + {\cal O}(\alpha_0)^2 \\&
+ \sum_{\alpha} \overline{(O_{\alpha\alpha})^2} p_{\alpha_0}(\alpha)^2 - 2{\cal O}(\alpha_0)\sum_\alpha \overline{O_{\alpha\alpha}}p_{\alpha_0}(\alpha) \\& 
= \sum_\alpha 
\left( \overline{\left(O_{\alpha \alpha}\right)^2} \ - { \cal O}(\alpha)^2 \right) p_{\alpha_0}(\alpha)^2 \\& \qquad + \left( \sum_\alpha {\cal O}(\alpha) p_{\alpha_0}(\alpha) - {\cal O}(\alpha_0) \right)^2 \\&
:= \Delta^2_{O,1} +  \Delta^2_{O,2}
\end{split}
\end{equation}
where to arrive at the second equality we add and subtract the term 
$\sum_{\alpha} \overline{O_{\alpha \alpha}}^2 p(\alpha)_{\alpha_0}^2$, and further use that that $\overline{O_{\alpha\alpha}O_{\beta\beta}} = {\cal O}(\alpha){\cal O}(\beta)$ for $\alpha \neq \beta$.

Deviations from the approximation \eqref{eq:smoothness_p_alpha} therefore come from two terms: (i) $\Delta_{O,1}$, which depends on both the variance of $O_{\alpha\alpha}$, and $p_{\alpha_0}^2$. The variance of $O_{\alpha\alpha}$ will be bounded for spin operators by 1, whereas $\sum_\alpha p_{\alpha_0}(\alpha)^2$ is of order $\frac{1}{\gamma_p D_{\alpha_0}}$, where $D_{\alpha_0}$ is the DOS at the peak of the distribution $p_{\alpha_0}(\alpha)$. (ii) $\Delta_{O,2}$, which assuming that ${\cal O}(\alpha)$ is almost constant within an interval $\gamma_p$, can be approximated around $\alpha_0$ in the form of a Taylor series, 
${\cal O}(\alpha) \approx 
{\cal O}(\alpha_0) + {\cal O}'(\alpha_0)(\alpha-\alpha_0) + (1/2){\cal O}''(\alpha_0)(\alpha-\alpha_0)^2$, such that
\begin{equation}
\begin{split}
\Delta^2_{O,2} & \approx {\cal O}(\alpha_0)\sum_\alpha p_{\alpha_0}(\alpha) + {\cal O}^\prime(\alpha_0)\sum_\alpha p_{\alpha_0}(\alpha)(\alpha_0 - \alpha) \\&
+{\cal O}^{\prime\prime}(\alpha_0)\sum_\alpha p_{\alpha_0}(\alpha)(\alpha_0 - \alpha)^2 - {\cal O}(\alpha_0) \\&
= \frac{1}{4}{\cal O}^{\prime\prime}(\alpha_0)^2\gamma_p^4.
\end{split}
\end{equation}
Therefore, we see that $\Delta_{O,2}$ is simply the variation in $\cal O$ over the width $\gamma_p$. Thus, we recover the conditions of Eq. \eqref{smoothness}.

These considerations thus validate our intuition that, as long as the mean value of $O_{\alpha \alpha}$ varies slowly with respect to $\alpha$, matrix elements $O_{\alpha \alpha}$ can be substituted by their average within summations over a large enough number of states in the non-interacting basis.


The smoothness condition is very reasonable when considered together with the sparsity condition above. Consider the product state basis defined in \eqref{product.basis}. A local observable can be written as $O = O_S \otimes \mathbb{1}_B$. Diagonal matrix elements, for example, are given by
\begin{equation}
O_{\alpha \alpha} = (O_S)_{s(\alpha),s(\alpha)} \delta_{\alpha_B(\alpha),\alpha_B(\alpha)},
\end{equation}
which implies that these matrix elements of the local operator $O_{S}$ are distributed along the diagonal of $O$, in an order that will be determined by the energy ordering of states $|\phi_{\alpha}\rangle$.

\subsection{Eigenstate Thermalization Hypothesis}

The assumptions on observables detailed above may be exploited to derive both the diagonal, and off-diagonal parts of the ETH, the form of which is given by Srednicki's ansatz \cite{Srednicki1999}:
\begin{equation}\label{eq:ETH}
O_{\mu\nu} = \mathcal{O}(E) + \frac{1}{\sqrt{D(E)}}f(E, \omega)\mathcal{R}_{\mu\nu},
\end{equation}
where $\mathcal{O}(E)$ and $f(E, \omega)$  are smooth functions of their respective arguments, $E = \frac{E_\mu + E_\nu}{2}$ and $\omega = E_\mu - E_\nu$, $D(E)$ is the density of states, and $R_{\mu\nu}$ is a stochastic variable of mean zero and unit variance. Each term of the ETH is derived in Appendix \ref{App:ETH}, for observables satisfying sparsity and smoothness conditions. 

To describe the process of quantum thermalization consistently, both diagonal and off-diagonal elements of observables play important roles. We will see that the off-diagonal elements dictate both the route to equilibrium, as well as the time-averaged fluctuations, and are thus the main focus of our work. The diagonal elements, however, dictate the equilibrium value of a given observable, and are thus similarly indispensable for a consistent theory of thermalization. For the diagonal elements, our RMT predicts that 
\begin{equation}\label{eq:O_diags}
O_{\mu\mu} \approx \overline{[O_{\alpha\alpha}]}_\mu,
\end{equation}
where $\overline{[O_{\alpha\alpha}]}_\mu$ is given in Eq. \eqref{def.average} with $n = 0$, and
\begin{equation}\label{eq:Obs_dist}
|O_{\mu\nu}|^2_{\mu\neq\nu} \approx \sum_n a_n \Lambda^{(2)}(\mu, \nu - n),
\end{equation}
where we define
\begin{equation}
\Lambda^{(n)}(\mu, \nu) := \frac{\omega_0 n\Gamma / \pi}{(E_\mu - E_\nu)^2 + (n\Gamma)^2},
\end{equation}
and 
\begin{equation}\label{eq:a_n}
    a_n =
    \begin{cases*}
      \overline{[\Delta O_{\alpha\alpha}^2]}_{\overline{\mu}}  := \overline{[O^2_{\alpha\alpha}]}_{\overline{\mu}} - \overline{[O_{\alpha\alpha}]}_{\overline{\mu}}^2 & if $n = 0$ \\
      \overline{[O_{\alpha,\alpha+n}^2]}_{\tilde{\mu}}       & otherwise.
    \end{cases*}
  \end{equation}
Here the microcanonical averages of matrix elements are centred around $\overline{\mu} = (\mu + \nu)/2$, and $\tilde{\mu} = (\mu + \nu - n)/2$ respectively. 
We thus see that off-diagonal matrix elements $|O_{\mu\nu}|_{\mu\neq\nu}^2$ are described by Lorentzians of width $2\Gamma$ \cite{Nation2018}, with peaks at energies $E_n = \omega_0 n$ separating those states coupled by $O$. 

In Appendix \ref{App:ETH}, we further show that the form obtained for the diagonal elements, Eq. \eqref{eq:O_diags}, obtains the correct long-time average for observables. For the remainder of this work, we focus on the role of off-diagonal elements, which are the key aspect that determine both the route to equilibrium, and the fluctuations thereafter.

In Eqs. \eqref{eq:O_diags} and \eqref{eq:Obs_dist}, and in the rest of this work, we use "$\approx$" as an approximation that is valid to leading order in $\frac{\omega_0}{\Gamma}$. 

\begin{figure*}
  \includegraphics[width=0.95\linewidth]{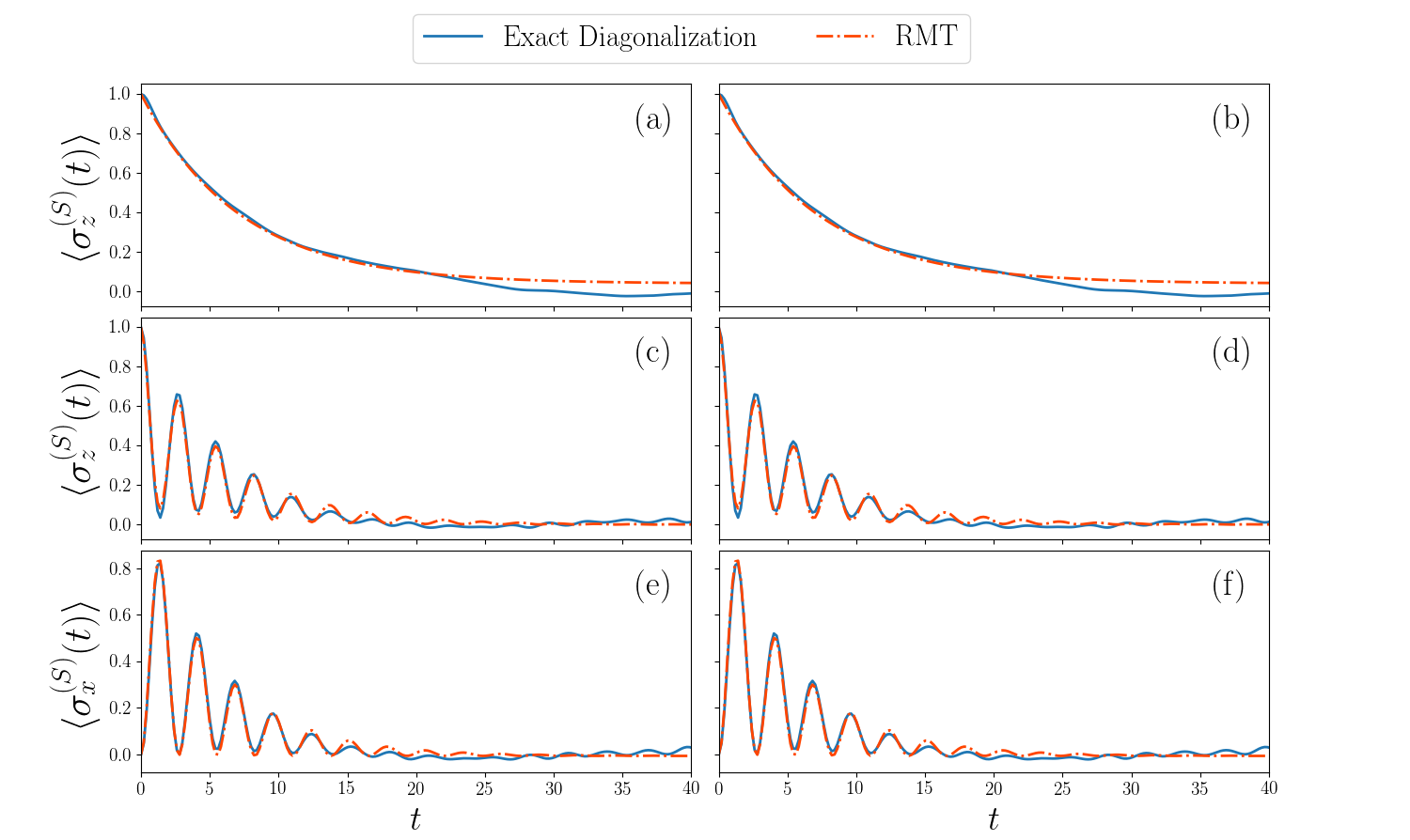}
 \caption{Time dependence of the Spin-Chain described be Eq. \eqref{eq:SC_Hamil} using Exact Diagonalization (blue line) for initial eigenstates of $H_0 = H_S + H_B$, $|\uparrow\rangle_S|\phi_\alpha\rangle_B$ [left column, (a), (c), (e)], and initial product states $|\uparrow\rangle_S|\uparrow, \downarrow, \cdots \rangle_B$ [right column, (b), (d), (f)]. System fields are $B_z^{(S)} = 0.8$, and $B_x^{(S)} = 0$ [top row, (a), (b)] and $B_x^{(S)} = 0.8$ [middle and bottom rows (c)-(f)]. Our RMT result of Eq. \eqref{eq:O_t} (red dot-dashed lines) is shown as a fit to obtain $\Gamma$. $\langle O(t) \rangle_0 = 1$ in (a), (b) is given by Eq. (\ref{eq:O_0t_fin}) in (c), (d) and by the analogous expression to that in (\ref{eq:O_0t_fin}) for $\sigma_x^{(S)}$ in (e), (f). Other parameters used are $N = 13, J_x^{(SB)} = 0.4, J_x = 1, B_z^{(B)} = 0, B_x^{(B)} = 0.3, J_z^{(SB)} = 0.2, J_z = 0$.
  }
  \label{fig:Time_Dep}
\end{figure*}
\section{Time-Dependence of Observables}
From the details outlined above, we are now able to derive the full time dependence of observables satisfying our physical conditions. We will further see that important features of thermalization may be observed even without appeal to our conditions on observables, but are rather more generic. A full account of the dynamics of thermalization is revealed by application of the sparsity and smoothness assumptions of Section \ref{Sec:Assumptions}, as well as the self-averaging property of random matrices, Eq. \eqref{eq:self_averaging}.

We consider the time evolution of an observable $O$, starting from an arbitrary initial pure state, 
\begin{equation}
|\psi(0)\rangle = \sum_{\alpha_0}\psi_{\alpha_0}|\phi_{\alpha_0}\rangle,
\end{equation}
where $\{|\phi_{\alpha_0}\rangle\}$ labels the basis of eigenstates of the non-interacting Hamiltonian $H_0$. 
We begin by defining the quantity
\begin{equation}
\Delta O (t) := \langle O(t) \rangle - \overline{\langle O(t) \rangle},
\end{equation}
where
\begin{equation}\label{eq:time_ave_O}
\overline{\langle O(t) \rangle} := \lim_{T\to \infty} \frac{1}{T}\int_0^T dt \langle O(t) \rangle .
\end{equation}

We may then write, assuming that the energies $E_\mu$ are non-degenerate,
\begin{equation}
\begin{split}
\Delta O(t) = \sum_{\substack{\alpha_0, \beta_0, \\ \alpha, \beta}}\sum_{\substack{\mu, \nu \\ \mu \neq \nu}} & \psi_{\alpha_0}\psi^*_{\beta_0}c_\mu(\alpha_0)c_\nu(\beta_0)c_\mu(\alpha)c_\nu(\beta) \\& \times O_{\alpha\beta}e^{-i(E_\mu - E_\nu)t}.
\end{split}
\end{equation}
Now, assuming self averaging, we treat the observable as equal to its ensemble average, such that $\Delta O(t) = \langle \Delta O(t) \rangle_V$. We then find
\begin{equation}\label{eq:DeltaO}
\begin{split}
\Delta O(t) = \sum_{\substack{\alpha_0, \beta_0, \\ \alpha, \beta}}\sum_{\substack{\mu, \nu \\ \mu \neq \nu}} &  \psi_{\alpha_0}\psi^*_{\beta_0}\langle c_\mu(\alpha_0)c_\nu(\beta_0)c_\mu(\alpha)c_\nu(\beta)\rangle_V  \\& \times O_{\alpha\beta}e^{-i(E_\mu - E_\nu)t}.
\end{split}
\end{equation}
We thus observe that the time evolution may be written in terms of the four-point correlation function $\langle c_\mu(\alpha_0)c_\nu(\beta_0)c_\mu(\alpha)c_\nu(\beta)\rangle_V$ of the off-diagonal ($\mu\neq\nu$) terms only. This correlation function was found in Ref. \cite{Nation2018}, and is given in Appendix \ref{App:Summary}, Eq. (\ref{eq:Corr_Func}). Substituting this into Eq. \eqref{eq:DeltaO}, we have,
\begin{equation}\label{eq:O_arb_full}
\begin{split}
\Delta & O(t) = \sum_{\substack{\mu, \nu \\ \mu \neq \nu}} \bigg[ \sum_{\alpha_0, \beta_0} \psi_{\alpha_0}\psi^*_{\beta_0} O_{\alpha_0\beta_0} \Lambda(\mu, \alpha_0)\Lambda(\nu, \beta_0) \\& - \sum_{\alpha_0, \alpha} |\psi_{\alpha_0}|^2 O_{\alpha\alpha}\frac{\Lambda(\mu, \alpha_0)\Lambda(\nu, \alpha_0)\Lambda(\mu, \alpha)\Lambda(\nu, \alpha)}{\Lambda^{(2)}(\mu, \nu)}  \\ & - \sum_{\alpha_0, \beta_0}\psi_{\alpha_0}\psi^*_{\beta_0} O_{\alpha_0\beta_0}  \frac{\Lambda(\mu, \alpha_0)\Lambda(\nu, \alpha_0)\Lambda(\mu, \beta_0)\Lambda(\nu, \beta_0)}{\Lambda^{(2)}(\mu, \nu)}\bigg] \\& \qquad \qquad \times e^{-i(E_\mu - E_\nu)t}.
\end{split}
\end{equation}

Now, noting that for the bulk states we analyze we have $\Lambda(\mu, \alpha) = \Lambda(\mu - \alpha)$, we may evaluate the first term in \eqref{eq:O_arb_full} by defining $\tilde{\mu} = \mu - \alpha_0$, $\tilde{\nu} = \nu - \beta_0$, and thereby obtain
\begin{equation}\label{eq:initial_state_decay}
\begin{split}
 \sum_{\alpha_0, \beta_0} \psi_{\alpha_0}\psi^*_{\beta_0} O_{\alpha_0\beta_0} & e^{-i(E_{\alpha_0} - E_{\beta_0} )t} \sum_{\substack{\tilde{\mu}, \tilde{\nu}}}  \Lambda(\tilde{\mu})\Lambda(\tilde{\nu}) e^{-i(E_{\tilde{\mu}} - E_{\tilde{\nu}} )t} \\& = \langle O(t) \rangle_0 e^{-2\Gamma t},
 \end{split}
\end{equation}
where $\langle O(t) \rangle_0$ is the evolution of the observable $O$ under the non-interacting Hamiltonian $H_0$, and we have taken the continuum limit of the summation $\sum_{\tilde{\mu}} \to \int\frac{dE_{\tilde{\mu}}}{\omega_0}$, such that we obtain Fourier transforms of each $\Lambda$, which results in the exponentially decaying factor. 

We stress here that Eq. \eqref{eq:initial_state_decay} did not require any assumption on the observable $O$, only the self-averaging property. We comment further on the implications of this at the end of this section.

Now, to evaluate the second term in \eqref{eq:O_arb_full} we require the smoothness condition (see Section \ref{Sec:Assumptions}). Explicitly, applied here, this can be seen as the removal of the microcanonical average of matrix elements from a summation of the form $\sum_\alpha O_{\alpha\alpha} \Lambda(\mu, \alpha)\Lambda(\nu, \alpha) \approx \overline{[O_{\alpha\alpha}]}_{\overline{\mu}}\Lambda^{(2)}(\mu, \nu)$, with $\overline{\mu} = \frac{\mu + \nu}{2}$. We thus see that the second term in \eqref{eq:O_arb_full} is given by
\begin{equation}\label{eq:selection}
\begin{split}
 & \sum_{\substack{\mu,\nu\\ \mu \neq \nu}}  \frac{\sum_{\alpha_0} |\psi_{\alpha_0}|^2\Lambda(\mu, \alpha_0)\Lambda(\nu, \alpha_0)\sum_\alpha O_{\alpha\alpha}\Lambda(\mu, \alpha)\Lambda(\nu, \alpha)}{\Lambda^{(2)}(\mu, \nu)} \\ & \qquad \qquad \qquad  \qquad \qquad  \qquad \times e^{-i(E_\mu - E_\nu)t} \\ &= 
\sum_{\substack{\mu,\nu\\ \mu \neq \nu}} \sum_{\alpha_0}\overline{[O_{\alpha\alpha}]}_{\overline{\mu}} |\psi_{\alpha_0}|^2 \Lambda(\mu - \alpha_0)\Lambda(\nu-\alpha_0)  e^{-i(E_\mu - E_\nu)t} \\& \approx \overline{[O_{\alpha\alpha}]}_{\overline{\alpha_0}}\sum_{\substack{\tilde{\mu}, \tilde{\nu}\\ \tilde{\mu} \neq \tilde{\nu}}}\sum_{\alpha_0} |\psi_{\alpha_0}|^2 \Lambda(\tilde{\mu})\Lambda(\tilde{\nu})  e^{-i(E_{\tilde{\mu}} - E_{\tilde{\nu}})t} \\& = \overline{[O_{\alpha\alpha}]}_{\overline{\alpha_0}} e^{-2\Gamma t},
\end{split}
\end{equation}
where we have defined $\tilde{\mu} = \mu - \alpha, \, \tilde{\nu} = \nu - \alpha$, and $\overline{\alpha_0}$ is the central energy of the distribution $\{\psi_{\alpha_0}\}$. For the third step we have used $\sum_{\alpha_0} \overline{[O_{\alpha\alpha}]}_{\alpha_0}|\psi_{\alpha_0}|^2 = \overline{[O_{\alpha\alpha}]}_{\overline{\alpha_0}}\sum_{\alpha_0} |\psi_{\alpha_0}|^2 =  \overline{[O_{\alpha\alpha}]}_{\overline{\alpha_0}}$, which can be seen to be a straightforward application of Eq. \eqref{smoothness.assumption}, and requires that the average $\overline{[O_{\alpha\alpha}]}_{\overline{\alpha_0}}$ is approximately constant over the width of the initial state distribution $\{\psi_{\alpha_0}\}$.

The third term in Eq. \eqref{eq:O_arb_full} is shown in Appendix \ref{App:Bound} to be bounded for all time by $\max_{\alpha_0\beta_0}(O_{\alpha_0\beta_0}) N_O\frac{3\omega_0 }{4\Gamma}$, which is small in comparison to other terms in the time evolution, and can thus be ignored. We note here that the sparsity condition is required in order to arrive at this bound.
  
For the time evolution of generic observables, we thus obtain
\begin{equation}\label{eq:O_t1}
\langle O(t) \rangle = \bigg(\langle O(t) \rangle_0 - \overline{[O_{\alpha\alpha}]}_{\overline{\alpha_0}} \bigg) e^{-2\Gamma t} + \overline{\langle O(t) \rangle} + \mathcal{O}\left(\frac{\omega_0}{\Gamma}\right).
\end{equation}
Interestingly, from the conditions $\langle O(t=0) \rangle = \langle O(t=0) \rangle_0$, Eq. \eqref{eq:O_t1} requires that the microcanonical average around the initial state energy $\overline{[O_{\alpha\alpha}]}_{\overline{\alpha_0}}$ is equal to the time average $\overline{\langle O(t) \rangle}$ up to an error on the order $\mathcal{O}(\frac{\omega_0}{\Gamma})$. We note that this long-time value can also be derived from the diagonal correlation function, Eq. \eqref{eq:4pt_diag}, which is shown in Appendix \ref{App:ETH}. Thus, the dominating contribution becomes
\begin{equation}\label{eq:O_t}
\langle O(t) \rangle \approx \langle O(t) \rangle_0 e^{-2\Gamma t} + \overline{\langle O(t) \rangle}(1 - e^{-2\Gamma t}).
\end{equation}
This is the main analytic result of this work. We note that the form is particularly useful, as for most systems of interest obtaining $\langle O(t) \rangle_0$ is a trivial calculation, as it characterises the time evolution in the non-interacting Hamiltonian. We further note that a statistical theory for random wavefunctions $c_\mu(\alpha)$ that includes correlations induced by the orthonormality constraint is strictly required to arrive to Eq. \eqref{eq:O_t}.

For our applications below, Eq. \eqref{eq:O_t} provides a method of obtaining $\Gamma$ from the observable time dependence via a fit, which may account for non-trivial free evolution of the observable caused by \emph{e.g}, a magnetic field. A specific application to such a case is shown in Appendix \ref{App:Obs_x_field}, and its time-dependence shown in Fig. (\ref{fig:Time_Dep}). 

As $\Gamma$ is the width of the random wavefunctions, and thus of the local density of states (LDOS), it may thus be obtained by a fit to the time-dependence of the survival probability, which is in general challenging for a many-body system. Eq. \eqref{eq:O_t} may be seen as an extension of this methodology to generic observables. We will see below that combined with previous results on the time-averaged observable fluctuations \cite{Nation2018} (see Appendix \ref{App:Generic_QCFDT} for details and extension of previous results), Eq. \eqref{eq:O_t} provides an experimental protocol to test the applicability of the random matrix approach to realistic systems, as well as a method of measuring their DOS, in the form of an emergent classical FDT.

We further comment on some details of this derivation, and the form of Eq. \eqref{eq:O_t}, that provide some insight into the implications of our assumptions. As noted above, the first term in Eq. \eqref{eq:O_t} is obtained without the need for any assumptions on the observable $O$, only requiring that the system is self-averaging. We can see that this term is, in essence, a `decay of the initial observable value'. The second term in Eq. \eqref{eq:O_t}, which may be interpreted as a `grow-in of the microcanonical average', requires the smoothness assumption - namely, that a consistent microcanonical average may be defined over the width $\Gamma$. In Appendix \eqref{App:Bound}, we required the sparsity condition in order to show that the third term in Eq. \eqref{eq:O_t} may be neglected.

Indeed, then, a consistent theory of thermalization may be developed on the basis of (i) self-averaging, which dictates that information about the initial state decays in time; (ii) the ability to define a microcanonical average via the smoothness condition, which, intuitively, allows the system to decay to the microcanonical value; and (iii)  the sparsity constraint, which reduces the contribution of off-diagonal elements $O_{\alpha\beta}$ in the decay process, which then simply contribute through their effect on the free evolution $\langle O(t)\rangle_O$. 

Aside from the time-averaged observable expectation value being equal to the microcanonical average, a further requirement for thermalization is that the fluctuations around the equilibrium value are small. It is these fluctuations that are the focus of the remainder of this work, which we will see can be quantified analytically based on the same constraints.

\begin{figure}
\includegraphics[width=\linewidth]{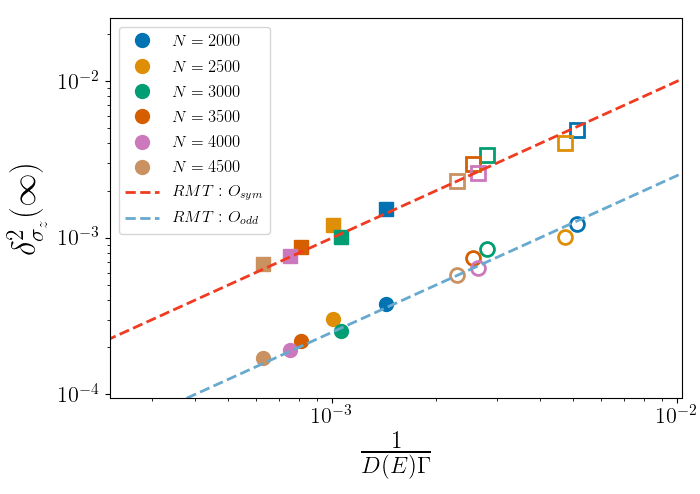}
  \caption{QC-FDT [Eq. (\ref{eq:FDT1})] for Random Matrix Hamiltonian. Squares show relation for $O_{\textrm{odd}}$, and diamonds for $O_{\textrm{sym}}$. Filled markers represent $g = 0.1$, unfilled represent $g = 0.05$. $O_{\textrm{odd}}$ and $O_{\textrm{sym}}$ differing as $\overline{[\Delta O_{\alpha\alpha}^2]}_{\alpha_0}$ is equal to $1/4, 1$ for $O_{\textrm{odd}}$ and $O_{\textrm{sym}}$, respectively. No averaging over realizations of the random matrix $V$ is performed, and hence we observe the self-averaging property.}
  \label{fig:RM}
\end{figure}

\section{Fluctuations from RMT}%
We now focus on the time-averaged fluctuations of an observable $O$, defined by
\begin{equation}\label{eq:flucs_def}
\begin{split}
\delta_O^2(\infty):= \lim_{T\to\infty} \Bigg[  &\frac{1}{T} \int_0^T dt \langle O(t) \rangle^2 \\ &- \left(\frac{1}{T}\int_0^T dt \langle O(t) \rangle\right)^2 \Bigg].
\end{split}
\end{equation}
Let us assume for now that the system is initially in an eigenstate of $H_0$, $|\phi_{\alpha_0}\rangle$, with energy $E_{\alpha_0}$.
The off-diagonal elements $O_{\mu\nu}$ govern the infinite-time fluctuations of $O$ \cite{DAlessio2016}, via,
\begin{equation}\label{eq:DE_fluc}
\delta_O^2(\infty) = \sum_{\substack{\mu, \nu \\ \mu \neq \nu}}|c_\mu(\alpha_0)|^2 |c_\nu(\alpha_0)|^2|O_{\mu\nu}|^2,
\end{equation}
where we have assumed that the energies $E_\mu$ are non-degenerate. In order to evaluate Eq. \eqref{eq:DE_fluc}, we may `decouple' the coefficients $c_\mu(\alpha)$ describing the initial state part (with subscripted indices $\alpha_0$), and observable, in the sense that, after performing the self-averaging assumption, we can write,
\begin{equation}
\begin{split}
\langle |c_\mu(\alpha_0)|^2 |c_\nu(\alpha_0)|^2 & |O_{\mu\nu}|^2\rangle_V \to \\& \langle |c_\mu(\alpha_0)|^2 |c_\nu(\alpha_0)|^2\rangle_V\langle |O_{\mu\nu}|^2\rangle_V.
\end{split}
\end{equation}
This is shown in Appendix \ref{App:Flucs}. Then, following Ref. \cite{Nation2018}, using Eqs. (\ref{eq:Lambda}), (\ref{eq:Obs_dist}), and (\ref{eq:DE_fluc}), we may convert the summations to integrals by the prescription $\sum_{\mu} \to \int dE_\mu D(E_\mu) = \int \frac{dE_\mu}{\omega_0}$, where $D(E)$ is the DOS. 

For the simplest case where $O$ is diagonal in the non-interacting basis, such that $n = 0$, we obtain
\begin{equation}\label{eq:FDT1}
\delta_O ^2(\infty) \approx \frac{\omega_0}{4\pi\Gamma}\overline{[\Delta O_{\alpha\alpha}^2]}_{\alpha_0}.
\end{equation}
We note that the same relation holds up to a factor even if $\Lambda(\mu,\alpha)$ has another form, such as Gaussian \cite{Mondaini2017, Santos2012, Atlas2017}, which we would expect outside of the low coupling regime. 
Eq. \eqref{eq:FDT1} shows an inverse relation between the observable time-fluctuations, 
$\delta^2_O(\infty)$, 
and the decay rate, $\Gamma$. 
We hereby refer to this result as the Quantum Chaotic Fluctuation-Dissipation Theorem (QC-FDT), since it establishes an effective description of $O(t)$ in terms of an effective Ornstein-Uhlenbeck process. 
\begin{figure}
  \includegraphics[width=0.99\linewidth]{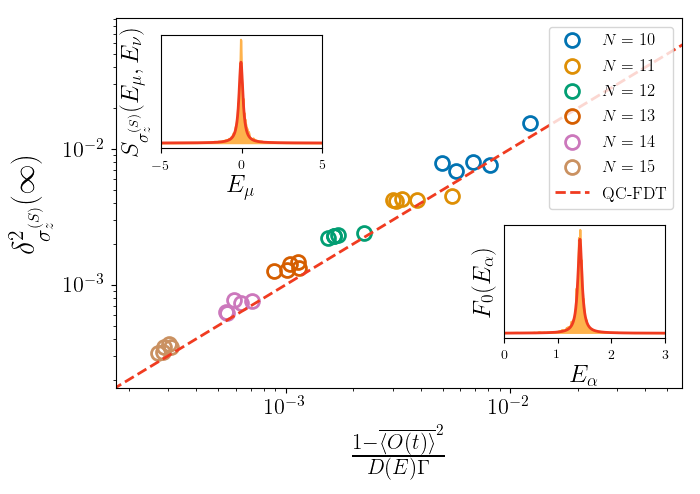}
  \caption{QC-FDT for Hamiltonian (\ref{eq:SC_Hamil}), observable $O = \sigma_z^{(S)}$. 
    Initial state given by $|\uparrow\rangle_S|\phi_\alpha\rangle_B$, where $H_B|\phi_\alpha\rangle_B = E^{(B)}_\alpha|\phi_\alpha\rangle_B$. Five different values of $\alpha$ are randomly selected from the central 1/2 of the energy spectrum $\{E_\alpha^{(B)}\}$. Parameters: $B_z^{(S)} = 0.8, \, B_z^{(B)} = 0, \, B_x^{(B)} = 0.3, \, J_z = 0.1, \, J_x = 1, \, J_x^{(SB)} = 0.4, \, J_z^{(SB)} = 0.2$.}
  \label{fig:SC}
\end{figure}

It has been previously observed numerically \cite{Borgonovi2017} that the fluctuations of observable matrix elements $O_{\mu\mu}$ (where fluctuations are defined by taking the average eigenstates close in energy to $|\psi_\mu\rangle$) decay as $1/N_{pc}$, where $N_{pc}$ is the number of principle components of a given eigenstate $|\psi_\mu\rangle$. We note that $N_{pc} \sim \Gamma D(E)$, and thus the QC-FDT shows this same relation.

In Fig. (\ref{fig:RM}) we present numerical results that demonstrate the QC-FDT for the RMT Hamiltonian (\ref{eq:Hamiltonian}). We obtain $\Gamma$ explicitly from a fit of the time dependence of the observables. 
The latter are given by $O_{odd}$ and $O_{sym}$, which are chosen to be diagonal in the non-interacting basis (thereby trivially fulfilling the sparsity condition), with diagonal elements,
\begin{equation}\label{eq:O_odd}
    (O_{odd})_{\alpha\alpha} =
    \begin{cases*}
      1 & if $\alpha = \textrm{odd}$ \\
      0 & otherwise,
    \end{cases*}
\end{equation}
for $O_{odd}$, and
\begin{equation}\label{eq:O_sym}
    (O_{sym})_{\alpha\alpha} =
    \begin{cases*}
      1 & if $\alpha = \textrm{odd}$ \\
      -1 & otherwise,
    \end{cases*}
\end{equation}
for $O_{sym}$.
These `observables' are chosen as they have a similar form to realistic observables made up of Pauli matrices: They are sparse, highly degenerate \cite{Anza2018}, and have a well defined structure in the non-interacting basis. These observables can further be seen to fulfil the smoothness conditions, as the average $\overline{[(O_{odd(sym)})_{\alpha\alpha}]}_{\alpha_0} = \frac{1}{2} (0)$ for all $\alpha_0$.

In a non-integrable quantum many-body system that is well described by our RMT model, we expect the QC-FDT \eqref{eq:FDT1} to hold, with the modification $\omega_0 \to 1/D(E_{\alpha_0})$, that is, we need to introduce the average energy level spacing at the initial energy $E_{\alpha_0}$.
\begin{figure}
  \includegraphics[width=0.99\linewidth]{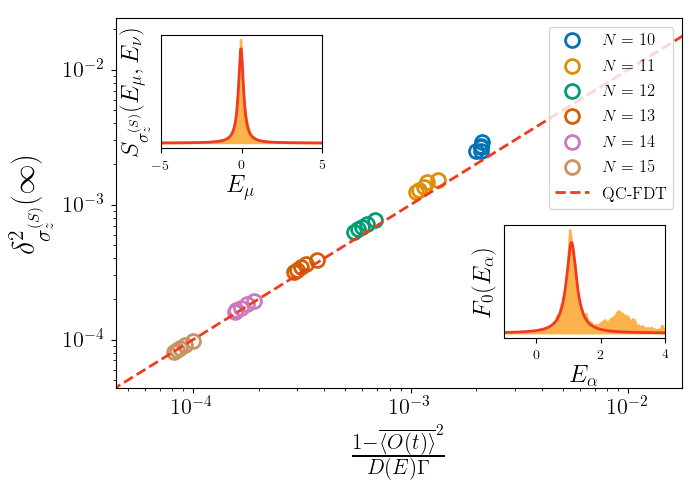}
  \caption{ QC-FDT  for Hamiltonian (\ref{eq:SC_Hamil}), observable $O = \sigma_z^{(S)}$. Initial state given by $|\psi(0)\rangle = |\uparrow\rangle_S|\downarrow, \downarrow, ..., \downarrow\rangle_B$.
   Parameters: $B_z^{(S)} = [0.4, 0.5, 0.6, 0.7, 0.8]$, all others equal to Fig. (\ref{fig:SC}).
  }
  \label{fig:SC2}
\end{figure}

\section{Numerics - Spin Chain Model}
We now investigate the applicability of the QC-FDT in quantum many-body Hamiltonians for the case described above, where $O_{\alpha\beta} \propto \delta_{\alpha\beta}$, and $|\psi(0)\rangle = |\phi_{\alpha_0}\rangle$, as described by Eq. \eqref{eq:FDT1}. Our model is a spin chain, with a Hamiltonian of the form,
\begin{equation}\label{eq:SC_Hamil}
H = H_S + H_B + H_{SB}.
\end{equation}
The system Hamiltonian $H_S$ describes a single spin in a $B_z$ field
\begin{equation}
H_S = B_{z}^{(S)}\sigma_z^{(1)},
\end{equation}
where $\{\sigma_i^{(j)}\}\quad i = {x, y, z}$ are the Pauli operators acting on site $j$. We take the system as site $j = 1$. The bath Hamiltonian is a spin-chain of length $N-1$, with nearest-neighbour Ising and XX interactions subjected to both $B_z$ and $B_x$ fields
\begin{equation}
\begin{split}
& H_B =  \sum_{j > 1}^N( B_{z}^{(B)}\sigma_z^{(j)}  + B_{x}^{(B)}\sigma_x^{(j)}) \\ & + \sum_{j > 1}^{N-1}[J_z\sigma_z^{(j)}\sigma_z^{(j+1)} + J_x(\sigma_+^{(j)}\sigma_-^{(j+1)} + \sigma_-^{(j)}\sigma_+^{(j+1)} )].
\end{split}
\end{equation}
The interaction part of the Hamiltonian describes a coupling of the system spin to a single bath ion of index $N_{\rm m}$,
\begin{equation}
\begin{split}
H_{SB} &= 
J_z^{(SB)}\sigma_z^{(1)}\sigma_z^{(N_{\rm m})} \\&+ J_x^{(SB)}(\sigma_+^{(1)}\sigma_-^{(N_{\rm m})} + \sigma_-^{(1)}\sigma_+^{(N_{\rm m})}),
\end{split}
\end{equation}
where $N_{\rm m} = 5$ throughout.
Thus we have $H_0 = H_S + H_B$, and $V = H_{SB}$. 

\begin{figure}
	\includegraphics[width=0.99\linewidth]{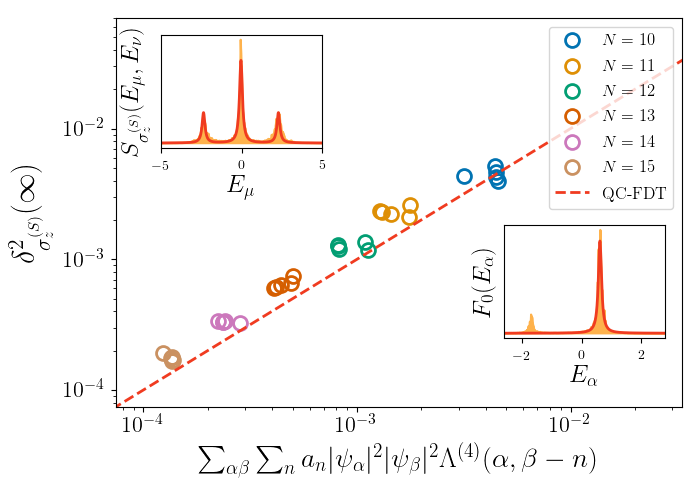}
	\caption{ Generalized QC-FDT  for Hamiltonian (\ref{eq:SC_Hamil}) with $H_S$ given by (\ref{eq:Bx_Hamil}). The QC-FDT is calculated explicitly for this case in Appendix \ref{App:QCFDT_BX}. Initial state given by $|\psi(0)\rangle = |\uparrow\rangle_S|\phi_\alpha\rangle_B$, where $\{|\phi_\alpha\rangle_B\}$ are the eigenstates of $H_B$.
	  Parameters: $B_z^{(S)} = B_x^{(S)} = 0.8$, all others equal to Fig. (\ref{fig:SC}).
	}
	\label{fig:Gen_FDT}
\end{figure}

\begin{figure}
  \includegraphics[width=0.99\linewidth]{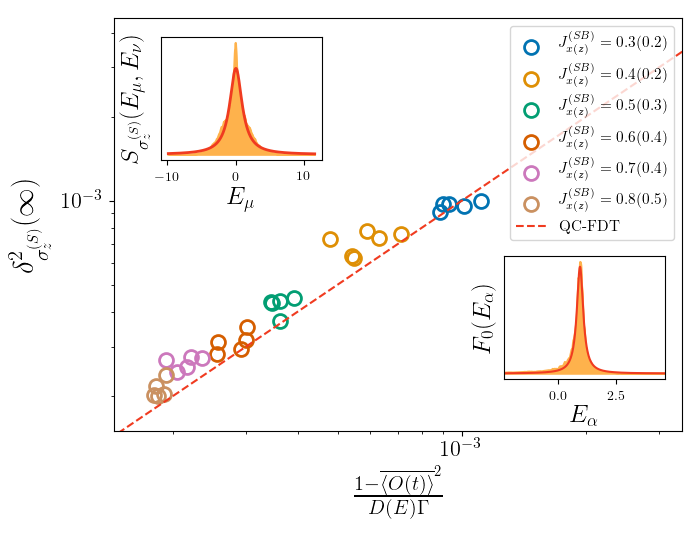}
  \caption{QC-FDT  for Hamiltonian (\ref{eq:SC_Hamil}) demonstrated by varying coupling strengths only, for initial states randomly selected from mid-energy eigenstates of $H_0$. Parameters: $J_x^{(SB)}$ and $J_z^{(SB)}$ shown in legend, $N = 14$, all others equal to Fig. (\ref{fig:SC}).
  }
  \label{fig:Coupling_FDT}
\end{figure}

In Fig. (\ref{fig:SC}) we present results for $N = 10, \dots,15$ and use as our observable $O = \sigma_z^{(1)}$.
In order to obtain $\Gamma$ we once again simulate the dynamics, and perform a fit to Eq. (\ref{eq:O_t}). 
We show the QC-FDT for initial states randomly selected from the set of states $\{|\uparrow\rangle_S|\phi_\alpha\rangle_B\}$, with $|\phi_\alpha\rangle_B$ denoting an eigenstate of $H_B$ with an energy in the central half of the spectrum $\{{}_B\langle\phi_\alpha|H_B|\phi_\alpha\rangle_B\}$. The insets of Figs. (\ref{fig:SC}), (\ref{fig:SC2}), (\ref{fig:Gen_FDT}), and (\ref{fig:Coupling_FDT}) show the smoothed initial state (bottom right) and observable (top left) distributions, obtained by the procedures 
\begin{equation}\label{eq:strength_function}
F_0(E_\alpha) = \sum_{\mu}|\langle\psi_\mu | \psi(0)\rangle |^2\delta_\epsilon(E_\mu - E_\alpha),
\end{equation}
for the initial state, and 
\begin{equation}
S_O(E_\mu, E_\nu) = \sum_{\mu\neq\nu}|O_{\mu\nu}|^2\delta_\epsilon(E_\mu - E_\nu),
\end{equation}
for observables, where $\delta_\epsilon(E_\mu - E) = \epsilon \pi^{-1} /[(E_\mu - E)^2 + \epsilon^2]$. Fits to Eqs. \eqref{eq:Lambda} and \eqref{eq:Obs_dist} are also shown (red line). We see that in each case we have a close agreement to a Lorentzian distribution, as expected from RMT.

\section{Generalized QC-FDT} 
So far we have focussed on the simplest case of the QC-FDT, namely, for observables $O_{\alpha\beta} \propto \delta_{\alpha\beta}$, and initial states $|\psi(0)\rangle = |\phi_{\alpha_0}\rangle$. 
In Appendix \ref{App:Generic_QCFDT} we extend this to all observables fulfilling both the sparsity and smoothness conditions, and arbitrary initial states $|\psi(0)\rangle = \sum_{\alpha}\psi_\alpha|\phi_\alpha\rangle$, assuming only that the smoothness condition may be applied over the distribution $\{\psi_\alpha\}$, as well as $\{c_\mu(\alpha)\}$.

For this, more general case, the time-averaged fluctuations are now described by
\begin{equation}\label{eq:flucs_gen}
\begin{split}
\delta_O^2(\infty) =\sum_{\alpha\beta\alpha^\prime\beta^\prime}&\sum_{\substack{\mu\nu\\ \mu\neq\nu}} \psi_\alpha\psi_\beta\psi_{\alpha^\prime}\psi_{\beta^\prime}  \\ & \times c_\mu(\alpha)c_\nu(\beta)c_\mu(\alpha^\prime)c_\nu(\beta^\prime)|O_{\mu\nu}|^2.
\end{split}
\end{equation}
As with the case above, see Eq. \eqref{eq:DE_fluc} and the following discussion, we observe that the correlations between coefficients of the initial state and observable decouple (see Appendix \ref{App:Flucs}), such that after taking the ensemble average $\langle \cdots \rangle_V$, we may substitute  
\begin{equation}\label{eq:decoupling_main}
\begin{split}
\langle c_\mu(\alpha)&c_\nu(\beta)c_\mu(\alpha^\prime)c_\nu(\alpha^\prime) |O_{\mu\nu}|^2 \rangle_V \to \\& \langle c_\mu(\alpha)c_\nu(\beta)c_\mu(\alpha^\prime)c_\nu(\alpha^\prime) \rangle_V \langle |O_{\mu\nu}|^2 \rangle_V,
\end{split}
\end{equation}
in Eq. \eqref{eq:flucs_gen}. 
Applying this, we obtain the following form for the generalized QC-FDT,
\begin{equation}\label{eq:FDT_General}
\delta_O^2(\infty) \approx \sum_{\alpha\beta}\sum_n a_n |\psi_\alpha|^2|\psi_\beta|^2 \Lambda^{(4)}(\alpha, \beta - n),
\end{equation}
which is shown in Appendix \ref{App:Generic_QCFDT} in detail.

We start our numerical analysis by applying Eq. \eqref{eq:FDT_General} to the case with an observable that is diagonal in the $H_0$ basis ($a_n = 0$ if $n \neq 0$).
In this case we can see that, as long as the energy width of $\psi_{\alpha}$ is much smaller than the decay rate $\Gamma$, we recover Eq. \eqref{eq:FDT1}. This is shown in Appendix \ref{App:Generic_QCFDT}, along with various examples of why we expect the simple form of the QC-FDT, Eq. \eqref{eq:FDT1}, to remain valid for many physical initial states. 

We test the QC-FDT numerically in this case by choosing a product state as an initial state  $|\psi(0)\rangle = |\uparrow\rangle_S|\downarrow, \downarrow, ..., \downarrow\rangle_B$.
This is shown in Fig. (\ref{fig:SC2}), where we see the same scaling predicted by Eq. \eqref{eq:FDT1}.

We have also numerically checked Eq. \eqref{eq:FDT_General} in the case in which the system observable $O$ is not diagonal in the basis of $H_0$, see Fig. (\ref{fig:Gen_FDT}). This case can be explored in our spin chain by adding an $x$-component to the system magnetic field, such that $H_S$ now reads
\begin{equation}\label{eq:Bx_Hamil}
H_S = B_{z}^{(S)}\sigma_z^{(1)} + B_{x}^{(S)}\sigma_x^{(1)}.
\end{equation}
In this case, the initial state $|\uparrow\rangle_S$ is no longer an eigenstate of $H_S$, and is instead given by a superposition $|\uparrow\rangle_S = \psi_+ |\phi_+\rangle_S + \psi_- |\phi_-\rangle_S$. The observable distribution $(|\sigma_z|^2)_{\mu\nu}$ is split into three peaks, located at $E_n = 0, \pm 2E$, where $E = \sqrt{( B_{x}^{(S)})^2+( B_{z}^{(S)})^2}$. We select the initial state of the bath to be a random mid energy eigenstate of $H_B$. We note in this case the approximation that the DOS does not change over relevant energy scales is a limiting factor, and may cause a deviation by a constant from the scaling seen in Eq. (\ref{eq:FDT_General}) for $E_n \gtrsim W$, where $W$ is the width over which the significant change in the DOS occurs. We calculate explicitly the form of the QC-FDT for this case, which is shown as the dashed line in Fig. \eqref{fig:Gen_FDT}, in Appendix \ref{App:QCFDT_BX}.

\section{Experimental Application} %
Finally, we discuss the possibility of an experimental observation of the QC-FDT. Ideally, we would like to test our result without the need of an exact numerical diagonalization of the closed quantum system. Both $\Gamma$ 
and $\delta^2_O(\infty)$ can be measured. However, the calculation of the DOS can be numerically challenging. One way around this problem is to calculate $D(E)$ for a non-interacting or integrable Hamiltonian that is sufficiently close to the real Hamiltonian. However, this approach relies on a detailed knowledge of the system and bath, and it may not always be possible.

A different approach is to explore the QC-FDT experimentally is to measure $\delta^2_O(\infty)$ and $\Gamma$ for a constant system size $N$ {\it but varying the coupling strength}. That is, assuming $V \propto g$, one could test the linear relation between $\delta^2_O(\infty)$ and $1/\Gamma$. We have numerically tested this approach as shown in Fig. (\ref{fig:Coupling_FDT}). Our ideas could indeed be used to characterize the dimension of quantum system in terms of the quantity $\delta^2_O(\infty) \Gamma$, which on average is proportional to the DOS that are participating in the quantum thermalization process.

\section{Conclusion and Outlook} 
In summary, we have obtained an analytic expression for the full time-dependence of the thermalization of physical observables to their microcanonical average. We further obtain an expression for the time averaged fluctuations of observables in chaotic quantum systems in terms of the rate of decay to equilibrium after a perturbation. 
Our results show the emergence of a classical fluctuation-dissipation relation, corresponding to an effective Ornstein-Uhlenbeck process, in a closed chaotic quantum system.
Our results rely on a RMT description of a quantum thermalization process in which an interaction term coupling two parts of the quantum system is suddenly switched on triggering a quantum thermalization process. 
In our approach the system-bath coupling is approximated by a Gaussian random matrix, an assumption that can be justified for a generic non-integrable system and weak system-bath couplings. We have successfully tested our result in a numerical experiment in a quantum spin chain.

Our result will help bridge the gap \cite{Merali2017} between theoretical results on quantum thermalization and experiments with closed quantum systems. 
In those cases in which a good approximation for the DOS can be calculated, a check of the QC-FDT would involve measurable quantities such as the decay rate and the time-fluctuations. Otherwise, the QC-FDT relation can still be checked experimentally as long as the coupling strength can be varied while keeping a constant system size. Our theory can thus be verified in quantum simulators working beyond the numerically tractable regime. Furthermore we argue that the product $\Gamma \delta^2_O(\infty)$ can indeed be considered as a measurement of the DOS of a non-integrable quantum system. As such, our work may prove useful in estimating the size of the Hilbert space in quantum devices.

We acknowledge funding by the People Programme (Marie Curie Actions) of the EU’s Seventh Framework Programme under REA Grant Agreement No. PCIG14-GA-2013-630955, and EPSRC grant no.
EP/M508172/1.\\

\appendix

\section{Summary of the RMT Approach}\label{App:Summary}

Below we present some necessary derivations for the results used in the main text. These are based on the random matrix formalism developed in reference \cite{Nation2018}, for the model used in the early work by Deutsch \cite{Deutsch1991, Deutscha}. We begin by summarizing the necessary results required for the following discussion, and refer the reader to reference \cite{Nation2018} for further details. 

In Ref. \cite{Nation2018} the current authors developed a consistent theoretical model of random wavefunctions $|\psi_\mu\rangle = \sum_\alpha c_\mu(\alpha)|\phi_\alpha\rangle$, for the  random matrix model described by Eq. (\ref{eq:Hamiltonian}). 
It is common in non-integrable systems and random matrix theory\cite{Deutsch1991, Reimann2015} to approximate the coefficients $c_\mu(\alpha)$ as Gaussian distributed random variables, however, it is shown in Ref. \cite{Nation2018} that this leads to inconsistent results for the off-diagonal matrix elements $O_{\mu\nu} := \langle\psi_\mu|O|\psi_\nu\rangle$ of observables, and also that the modification to account for orthogonality of eigenstates resolves this inconsistency. 

We thus modify the Gaussian probability distribution on the $c_\mu(\alpha)$s to require this orthogonality, using
\begin{equation}\label{eq:p_c}
p(c, \Lambda) = \frac{1}{Z_p}e^{-\sum_{\mu\alpha}\frac{c^2_\mu(\alpha)}{2\Lambda(\mu, \alpha)}}\prod_{\substack{\mu\nu\\ \mu>\nu}}\delta(\sum_{\alpha}c_\mu(\alpha)c_\nu(\alpha)),
\end{equation}
for some distribution $\Lambda(\mu, \alpha)$. This distribution was found to be a Lorentzian of width $\Gamma = \frac{\pi g^2}{N \omega_0}$ with no orthogonality condition in \cite{Deutscha}, and repeated for $p(c, \Lambda)$ above in Appendix A of \cite{Nation2018}. From Eq. (\ref{eq:p_c}), assuming that the dominant interactions are those of two eigenvectors only, one can calculate arbitrary correlation functions of the $c_\mu(\alpha)$ coefficient by first defining the generating function,
\begin{widetext}
\begin{equation}{\label{eq:gen_func_1}}
\begin{split}
G^{(\textrm{od})}_{\mu\nu} (\vec{\xi}_{\mu},\vec{\xi}_{\nu})& = \int \int \exp\bigg[-\sum_\alpha\bigg( \frac{c_\mu^2(\alpha)}{2\Lambda(\mu, \alpha)} + \frac{c_\nu^2(\alpha)}{2\Lambda(\nu, \alpha)} + \xi_{\mu,\alpha} c_{\mu}(\alpha) + \xi_{\nu,\alpha} c_{\nu}(\alpha) \bigg)\bigg] 
\delta(\sum_{\alpha}c_\mu(\alpha)c_\nu(\alpha))\prod_{\alpha}dc_\mu(\alpha) dc_\nu(\alpha) \\ &
\propto \exp \bigg[ \frac{1}{2} \sum_\alpha \xi^2_{\mu,\alpha} \Lambda(\mu,\alpha) + \frac{1}{2} \sum_\alpha \xi^2_{\nu,\alpha} \Lambda(\nu,\alpha) - 
\frac{1}{2} \sum_{\alpha,\beta} \xi_{\mu,\alpha} \xi_{\mu,\beta} \xi_{\nu,\alpha} \xi_{\nu,\beta}  \frac{\Lambda(\mu,\alpha) \Lambda(\mu,\beta) \Lambda(\nu,\alpha) \Lambda(\nu,\beta)}{\Lambda^{(2)}(\mu, \nu)}\bigg],
\end{split}
\end{equation}
\end{widetext}
where in the second line we have re-expressed the $\delta$-functions in their Fourier form. The superscript $(\textrm{od})$ indicates that this is the `off-diagonal' generating function, requiring $\mu \neq \nu$. The diagonal case is discussed below.
The correlation functions may then be calculated by performing successive derivatives with respect to the force terms $\xi$ via
\begin{equation}\label{eq:Derivs}
\begin{split}
\langle c_\mu&(\alpha)  c_\nu(\beta) \cdots c_\mu(\alpha_1^\prime) c_\nu(\beta_1^\prime) \rangle_V = \\& \frac{1}{G_{\mu\nu}} \partial_{\xi_{\mu,\alpha}} \partial_{\xi_{\nu,\beta}} \cdots \partial_{\xi_{\mu,\alpha_1^\prime}} \partial_{\xi_{\nu,\beta_1^\prime}} G_{\mu\nu} {\bigg |}_{\xi_{\mu,\alpha}=0,\xi_{\nu,\alpha}=0}.
\end{split}
\end{equation}
In particular, the correlation function $\langle c_\mu(\alpha_0)c_\nu(\beta_0)c_\mu(\alpha)c_\nu(\beta)\rangle_V$ was found in \cite{Nation2018} for $\mu\neq \nu$ to be equal to
\begin{equation}\label{eq:Corr_Func}
\begin{split}
\langle c_\mu& (\alpha_0)c_\nu(\beta_0) c_\mu(\alpha) c_\nu(\beta) \rangle_{V} = \Lambda(\mu, \alpha_0)\Lambda(\nu, \beta_0)\delta_{\alpha_0\alpha}\delta_{\beta_0\beta} \\& - \frac{\Lambda(\mu, \alpha_0)\Lambda(\nu, \alpha_0)\Lambda(\mu, \alpha)\Lambda(\nu, \alpha)\delta_{\alpha_0\beta_0}\delta_{\alpha\beta}}{\Lambda^{(2)}(\mu, \nu)}  \\ & - \frac{\Lambda(\mu, \alpha_0)\Lambda(\nu, \alpha_0)\Lambda(\mu, \beta_0)\Lambda(\nu, \beta_0)\delta_{\alpha_0\beta}\delta_{\beta_0\alpha}}{\Lambda^{(2)}(\mu, \nu)},
\end{split}
\end{equation}
for $\mu \neq \nu$, with
\begin{equation}
\Lambda^{(n)}(\mu, \nu) := \frac{\omega_0 n\Gamma / \pi}{(E_\mu - E_\nu)^2 + (n\Gamma)^2},
\end{equation}
where the superscript $(n)$ is left out for $n=1$. The latter two terms in Eq. (\ref{eq:Corr_Func}) arise as an explicit result of the orthogonality factor in Eq. (\ref{eq:p_c}). We comment further on the form of the correlation function (\ref{eq:Corr_Func}) at the beginning of Appendix \ref{App:Flucs}.

We stress here that the generating function Eq. \eqref{eq:gen_func_1} explicitly requires $\mu \neq \nu$, as it models the interactions due to mutual orthogonality of two random wavefunctions. For the diagonal part, we have the much simpler generating function,
\begin{equation}\label{eq:gen_func_diag}
G_{\mu\mu}^{(d)} = \int \exp\left[ -\sum_\alpha \frac{c_\mu^2(\alpha)}{2\Lambda(\mu, \alpha)}\right]\prod_\alpha dc_\mu(\alpha).
\end{equation}
Thus, we have,
\begin{equation}\label{eq:CF_diag}
\begin{split}
\langle c_\mu(\alpha)&c_\mu(\beta)c_\mu(\alpha^\prime)c_\mu(\beta^\prime)\rangle_V = \Lambda(\mu, \alpha)\Lambda(\mu, \alpha^\prime)\delta_{\alpha\beta}\delta_{\alpha^\prime\beta^\prime} \\& + \Lambda(\mu, \alpha)\Lambda(\mu, \beta)(\delta_{\alpha\alpha^\prime}\delta_{\beta\beta^\prime} + \delta_{\alpha\beta^\prime}\delta_{\alpha^\prime\beta}),
\end{split}
\end{equation}
for the diagonal case.

We note here that the generating functions above are general in the sense that they do not rely on any particular form of the distribution $\Lambda(\mu, \alpha)$. Indeed, for our model, with a homogeneous perturbation $V$, one can derive a Lorentzian form, see Eq. \eqref{eq:Lambda}, for the random-wavefunctions. As noted in the main text, one may expect in many situations for inhomogeneities in $V$ to become relevant. For example, in the case of local interactions and strong coupling, one expects the bandwidth $\Gamma_V$ to become relevant to the form of $\Lambda(\mu, \alpha)$. This would not, however, change the form of Eq. \eqref{eq:Corr_Func} or \eqref{eq:CF_diag}. In the case, then, where $\Lambda(\mu, \alpha)$ is described by a Gaussian, rather than a Lorentzian, which is common in spin-chain systems in the strong-coupling regime, one obtains a Gaussian decay, rather than exponential in Eq. \eqref{eq:O_t}, and a form of Eq. \eqref{eq:FDT1} that differs by a numerical prefactor \cite{Nation2018}.

\section{Full ETH from RMT}\label{App:ETH}

Here we calculate the diagonal, and off-diagonal matrix elements of observables from the above approach, using the sparsity and smoothness conditions outlined Section \ref{Sec:Assumptions}. 
\subsection{Diagonal ETH}
We can see that the diagonal matrix elements are given by
\begin{equation}
\begin{split}
\langle O_{\mu\mu} \rangle_V & = \sum_{\alpha\beta}\langle c_\mu(\alpha)c_\mu(\beta)\rangle_V O_{\alpha\beta} \\&
= \sum_\alpha \Lambda(\mu, \alpha) O_{\alpha\alpha} \\&
= \overline{[O_{\alpha\alpha}]}_\mu.
\end{split}
\end{equation}
One can observe that the fluctuations of the diagonal elements can also be analysed, considering the quantity
\begin{equation}\label{eq:O2_mumu}
\begin{split}
\langle O_{\mu\mu}^2 \rangle_V &= \sum_{\alpha\beta\alpha^\prime \beta^\prime}\langle c_\mu(\alpha)c_\mu(\beta)c_\mu(\alpha^\prime)c_\mu(\beta^\prime)\rangle_V O_{\alpha\beta}O_{\alpha^\prime\beta^\prime}\\&
=\sum_{\alpha\beta\alpha^\prime \beta^\prime}\bigg[\Lambda(\mu, \alpha)\Lambda(\mu, \alpha^\prime)\delta_{\alpha\beta}\delta_{\alpha^\prime\beta^\prime} \\& + \Lambda(\mu, \alpha)\Lambda(\mu, \beta)(\delta_{\alpha\alpha^\prime}\delta_{\beta\beta^\prime} + \delta_{\alpha\beta^\prime}\delta_{\alpha^\prime\beta})\bigg] O_{\alpha\beta}O_{\alpha^\prime\beta^\prime}\\& 
= \sum_{\alpha\beta}\Lambda(\mu,\alpha) \Lambda(\mu, \beta) \left( O_{\alpha\alpha}O_{\beta\beta} + O_{\alpha\beta}^2 + O_{\alpha\beta}O_{\beta\alpha} \right).
\end{split}
\end{equation} 
Now, we see that the first term in Eq. \eqref{eq:O2_mumu} is equal to $\langle O_{\mu\mu} \rangle_V^2$. For the second term, assuming the sparsity and smoothness conditions (see Section \ref{Sec:Assumptions}), we have
\begin{equation}
\begin{split}
\sum_{\alpha n} \Lambda(\mu, \alpha)\Lambda(\mu, \alpha + n)O_{\alpha\alpha+n}^2 & \approx \sum_n \overline{[O_{\alpha\alpha+n}^2]}_\mu\Lambda^{(2)}(\mu, n) \\&
\leq \sum_n \overline{[O_{\alpha\alpha+n}^2]}_\mu \frac{\omega_0}{2\pi\Gamma},
\end{split}
\end{equation}
and similarly, following the same approach we observe that the third term in Eq. \eqref{eq:O2_mumu} is bounded by $\sum_n \overline{[O_{\alpha+n\alpha}O_{\alpha\alpha+n}]}_\mu \frac{\omega_0}{2\pi\Gamma}$.
We thus observe that the fluctuations of the diagonal terms are small, in the sense that $\langle O_{\mu\mu}^2\rangle_V - \langle O_{\mu\mu} \rangle_V^2 \sim \mathcal{O}\left(\frac{\omega_0}{\Gamma}\right)$. 
Indeed, we can see that the smallness of the contributions of these terms is also necessary for the correct long-time average of observables Eq. \eqref{eq:time_ave_O}, which itself can be written as, for an arbitrary initial state $|\psi(0)\rangle = \sum_{\alpha_0}\psi_{\alpha_0}|\phi_{\alpha_0}\rangle$,
\begin{equation}
\overline{\langle O(t) \rangle} = \sum_\mu \sum_{\alpha_0\beta_0\alpha\beta}\psi_{\alpha_0}\psi^*_{\beta_0}c_\mu(\alpha_0)c_\mu(\beta_0)c_\mu(\alpha)c_\mu(\beta)O_{\alpha\beta}.
\end{equation}
Using Eq. \eqref{eq:CF_diag}, we have
\begin{equation}
\begin{split}
\overline{\langle O(t) \rangle}& = \sum_\mu \sum_{\alpha_0\beta_0\alpha\beta}\psi_{\alpha_0}\psi^*_{\beta_0}\bigg[\Lambda(\mu, \alpha_0)\Lambda(\mu, \alpha)\delta_{\alpha_0\beta_0}\delta_{\alpha_0\beta_0} \\& + \Lambda(\mu, \alpha_0)\Lambda(\mu, \beta_0)(\delta_{\alpha_0\alpha}\delta_{\beta_0\beta} + \delta_{\alpha_0\beta}\delta_{\alpha\beta_0})\bigg] O_{\alpha\beta}.
\end{split}.
\end{equation}
Now, we see that the latter two terms may be bounded by 
\begin{equation}
\sum_{\alpha_0\beta_0} \psi_{\alpha_0}\psi_{\beta_0}\Lambda^{(2)}(\alpha_0, \beta_0)O_{\alpha_0\beta_0} \leq \langle O(0)\rangle \frac{\omega_0}{2\pi\Gamma},
\end{equation}
which may be seen using that $\sum_\mu \Lambda(\mu, \alpha_0)\Lambda(\mu, \beta_0) = \Lambda^{(2)}(\alpha_0, \beta_0) \leq \frac{\omega_0}{2\pi\Gamma}$, and $\langle O(0) \rangle = \sum_{\alpha_0\beta_0}\psi_{\alpha_0}\psi_{\beta_0}O_{\alpha_0\beta_0}$. We thus see that these contributions are negligible in comparison to that of the first term:
\begin{equation}
\begin{split}
\overline{\langle O(t) \rangle} & = \sum_\mu \sum_{\alpha\beta} \Lambda(\mu, \alpha)\Lambda(\mu, \beta)|\psi_\alpha|^2 O_{\beta\beta}\\&
= \sum_\mu \overline{[\psi_\alpha]}_\mu \overline{[O_{\alpha\alpha}]}_\mu \\&
\approx O_{mc},
\end{split}
\end{equation}
which we can see returns the microcanonical average as required. We thus see that a consistent description of the long-time observable expectation value may be obtained in terms of our RMT approach. Moreover, we observe here that the microcanonical average of matrix elements described by the smoothness assumption emerges naturally as this equilibrium value.
\subsection{Off-diagonal ETH}
In order to calculate the distribution of the off-diagonal observable elements, we use the squared value (as they average to zero), and thus we write
\begin{equation}
|O_{\mu\nu}|^2 = \sum_{\alpha\beta\alpha^\prime\beta^\prime}c_\mu(\alpha)c_\nu(\beta)c_\mu(\alpha^\prime)c_\nu(\beta^\prime)O_{\alpha\beta}O_{\alpha^\prime\beta^\prime},
\end{equation}
which, assuming self-averaging, and using Eq. (\ref{eq:Corr_Func}), gives
\begin{equation}
\begin{split}
|O_{\mu\nu}|_{\mu\neq\nu}^2 & = \sum_{\alpha\beta}\Lambda(\mu, \alpha)\Lambda(\nu, \beta)O_{\alpha\beta}^2 \\& - \sum_{\alpha\alpha^\prime}\frac{\Lambda(\mu, \alpha)\Lambda(\nu, \alpha)\Lambda(\mu, \alpha^\prime)\Lambda(\nu, \alpha^\prime)}{\Lambda^{(2)}(\mu, \nu)} O_{\alpha\alpha}O_{\alpha^\prime\alpha^\prime}\\ & - \sum_{\alpha\beta}\frac{\Lambda(\mu, \alpha)\Lambda(\nu, \alpha)\Lambda(\mu, \beta)\Lambda(\nu, \beta)}{\Lambda^{(2)}(\mu, \nu)} O_{\alpha\beta}O_{\beta\alpha} .
\end{split}
\end{equation}
We separate this into terms describing diagonal, $O_{\alpha\alpha}$, and non-diagonal, $O_{\alpha\beta} \, | \, \alpha\neq\beta$, contributions,
\begin{widetext}
\begin{equation}
\begin{split}
|O_{\mu\nu}|_{\mu\neq\nu}^2 = \sum_{\alpha}\Lambda(\mu, \alpha) & \Lambda(\nu, \alpha)O_{\alpha\alpha}^2 - \sum_{\alpha\alpha^\prime}\frac{\Lambda(\mu, \alpha)\Lambda(\nu, \alpha)\Lambda(\mu, \alpha^\prime)\Lambda(\nu, \alpha^\prime)}{\Lambda^{(2)}(\mu, \nu)} O_{\alpha\alpha}O_{\alpha^\prime\alpha^\prime}(1 + \delta_{\alpha\alpha^\prime}) \\ &  + \sum_{\substack{\alpha\beta \\ \alpha\neq \beta}}\Lambda(\mu, \alpha)\Lambda(\nu, \beta)O_{\alpha\beta}^2 - \sum_{\substack{\alpha\beta \\ \alpha\neq \beta}}\frac{\Lambda(\mu, \alpha)\Lambda(\nu, \alpha)\Lambda(\mu, \beta)\Lambda(\nu, \beta)}{\Lambda^{(2)}(\mu, \nu)} O_{\alpha\beta}O_{\beta\alpha},
\end{split}
\end{equation}
\end{widetext}
and, as above, using the microcanonical averaging of matrix elements afforded by the smoothness assumption, $\sum_{\alpha}\Lambda(\mu, \alpha)\Lambda(\nu, \alpha)O_{\alpha\alpha} \approx \overline{[O_{\alpha\alpha}]}_{\overline{\mu}}\Lambda^{(2)}(\mu, \nu)$, on the diagonal contributions. We thereby obtain,
\begin{equation}\label{eq:O_sums}
\begin{split}
|O_{\mu\nu}|_{\mu\neq\nu}^2 & = \overline{[O_{\alpha\alpha}^2]}_{\overline{\mu}}\Lambda^{(2)}(\mu, \nu) - \overline{[O_{\alpha\alpha}]}_{\overline{\mu}}^2\Lambda^{(2)}(\mu, \nu)\\ & + \sum_{\substack{\alpha\beta \\ \alpha\neq \beta}}\Lambda(\mu, \alpha)\Lambda(\nu, \beta)O_{\alpha\beta}^2 \\& -  \sum_{\substack{\alpha\beta \\ \alpha\neq \beta}}\frac{\Lambda(\mu, \alpha)\Lambda(\nu, \alpha)\Lambda(\mu, \beta)\Lambda(\nu, \beta)}{\Lambda^{(2)}(\mu, \nu)} O_{\alpha\beta}O_{\beta\alpha},
\end{split}
\end{equation}
where the term in $\delta_{\alpha\alpha^\prime}$ is does not contribute, due to the reduced number of summations (see Appendix \ref{App:Flucs}). 
Now, to obtain the contribution of the latter two terms in Eq. \eqref{eq:O_sums}, we employ the sparsity assumption $O_{\alpha\beta} = \sum_{n \in N_O} O_{\alpha, \alpha+n}\delta_{\beta, \alpha+n}$, to obtain,
\begin{equation}
\begin{split}
 & |O_{\mu\nu}|_{\mu\neq\nu}^2 = \overline{[\Delta O_{\alpha\alpha}^2]}_{\overline{\mu}}\Lambda^{(2)}(\mu, \nu) \\& + \sum_{\alpha, n\neq 0}\Lambda(\mu, \alpha)\Lambda(\nu, \alpha + n)O_{\alpha, \alpha+n} O_{\alpha, \alpha + n}\\ & -  \sum_{\alpha, n\neq 0}\frac{\Lambda(\mu, \alpha)\Lambda(\nu, \alpha)\Lambda(\mu, \alpha + n)\Lambda(\nu, \alpha + n)}{\Lambda^{(2)}(\mu, \nu)} \\& \qquad \qquad \qquad \times O_{\alpha, \alpha + n}O_{\alpha + n, \alpha},
\end{split}
\end{equation}
where the summations over $n\neq0$ are understood to be on the set ${\cal N}_O$, as defined in Section \ref{Sec:Assumptions} as the off-diagonal finite matrix elements. Here we may see that the final term may be ignored, as the restricted summation relegates the order to $\sim \mathcal{O}\left(\left(\frac{\omega_0}{\Gamma}\right)^2\right)$. Finally, we may define an equivalent microcanonical averaging of matrix elements to that above for finite $n$, such that $\sum_{\alpha}\Lambda(\mu, \alpha)\Lambda(\nu, \alpha+n)O^2_{\alpha, \alpha+n} \approx \overline{[O_{\alpha, \alpha+n}^2]}_{\tilde{\mu}}\sum_{\alpha}\Lambda(\mu, \alpha)\Lambda(\nu, \alpha+n)$, where $\tilde{\mu} = \frac{\mu + \nu - n}{2}$, and $\overline{[O_{\alpha, \alpha+n}^2]}_{\tilde{\mu}} = \sum_{\alpha} \Lambda(\tilde{\mu}, \alpha)O^2_{\alpha, \alpha + n}$. We thus obtain
\begin{equation}
|O_{\mu\nu}|_{\mu\neq\nu}^2 = \overline{[\Delta O_{\alpha\alpha}^2]}_{\overline{\mu}}\Lambda^{(2)}(\mu, \nu) + \sum_{n\neq 0}\overline{[O_{\alpha, \alpha+n}^2]}_{\tilde{\mu}}\Lambda^{(2)}(\mu, \nu - n),
\end{equation}
which may be written as
\begin{equation}\label{eq:O_general}
|O_{\mu\nu}|_{\mu\neq\nu}^2 = \sum_n a_n \Lambda^{(2)}(\mu, \nu - n),
\end{equation}
where we have defined $a_n = a_n(E_{\overline{\mu}}) = \overline{[\Delta O_{\alpha\alpha}^2]}_{\overline{\mu}}$ for $n = 0$, and $\overline{[O_{\alpha, \alpha+n}^2]}_{\tilde{\mu}}$ otherwise. 

We thus observe that the square of the off-diagonal elements is given by a smooth function, proportional to $\omega_0 = \frac{1}{D(E)}$, which agrees with Srednicki's ansatz \cite{Srednicki1999}. We thus have that the full ETH is recovered from our RMT description. 

\section{A Bound}\label{App:Bound}

In this section we obtain a bound on the third term in Eq. \eqref{eq:O_arb_full}, and thus show that it is negligible in comparison to the others, which are obtained in the main text. The term we wish to bound is given by, $|A(t)|$, where
\begin{equation}
\begin{split}
A & (t) =\sum_{\substack{\mu,\nu\\ \mu \neq \nu}} \sum_{\alpha_0, \beta_0}  \psi_{\alpha_0}\psi^*_{\beta_0} O_{\alpha_0\beta_0} \\& \times \frac{\Lambda(\mu, \alpha_0)\Lambda(\nu, \alpha_0)\Lambda(\mu, \beta_0)\Lambda(\nu, \beta_0)}{\Lambda^{(2)}(\mu, \nu)} e^{-i(E_\mu - E_\nu)t}.
\end{split}
\end{equation}
We first note that no similar microcanonical averaging procedure to that used in the evaluation of the other terms in Eq. \eqref{eq:O_arb_full} can be performed, as the average would be required over the coefficients $\psi_\alpha$. This means that a sum over $\alpha_0$ or $\beta_0$ cannot be expected to cancel, even approximately, with the denominator. As such, the smoothness condition is not useful for this bound, which will be seen instead to be a feature of the sparsity local of observables in the non-interacting basis.

Now, we proceed using $|\sum_i a_i| \leq \sum_i |a_i|$ (which can be seen for any sequence $\{a_i\}$ by noting that the bound is saturated for when $a_i > 0 \; \forall \; i$, and that swapping the sign of any $a_i$ decreases the left-hand side, and the right-hand side remains the same), we can write
\begin{widetext}
\begin{equation}
\begin{split}
|A(t)| & \leq \sum_{\substack{\mu,\nu\\ \mu \neq \nu}}  \left| \sum_{\alpha_0, \beta_0} \psi_{\alpha_0}\psi^*_{\beta_0} O_{\alpha_0\beta_0}  \frac{\Lambda(\mu, \alpha_0)\Lambda(\nu, \alpha_0)\Lambda(\mu, \beta_0)\Lambda(\nu, \beta_0)}{\Lambda^{(2)}(\mu, \nu)} \right|\left| e^{-i(E_\mu - E_\nu)t} \right| \\ &
\leq \sum_{\substack{\mu,\nu\\ \mu \neq \nu}} \sum_{\alpha_0, \beta_0} \left| \psi_{\alpha_0}\psi^*_{\beta_0} O_{\alpha_0\beta_0}  \frac{\Lambda(\mu, \alpha_0)\Lambda(\nu, \alpha_0)\Lambda(\mu, \beta_0)\Lambda(\nu, \beta_0)}{\Lambda^{(2)}(\mu, \nu)} \right| \\ &
=\sum_{\alpha_0, \beta_0} \left| \psi_{\alpha_0}\psi^*_{\beta_0} O_{\alpha_0\beta_0} \right|  \sum_{\substack{\mu,\nu\\ \mu \neq \nu}} \frac{\Lambda(\mu, \alpha_0)\Lambda(\nu, \alpha_0)\Lambda(\mu, \beta_0)\Lambda(\nu, \beta_0)}{\Lambda^{(2)}(\mu, \nu)} \\ &
\leq \frac{3\omega_0}{4\pi\Gamma}\sum_{\alpha_0, \beta_0} \left| \psi_{\alpha_0}\psi^*_{\beta_0} O_{\alpha_0\beta_0} \right|,
\end{split}
\end{equation}
\end{widetext}
where we have used that 
\begin{equation}\label{eq:bound1}
\begin{split}
\sum_{\substack{\mu,\nu\\ \mu \neq \nu}} & \frac{\Lambda(\mu, \alpha_0)\Lambda(\nu, \alpha_0)\Lambda(\mu, \beta_0)\Lambda(\nu, \beta_0)}{\Lambda^{(2)}(\mu, \nu)} \\ & \qquad =  \omega_0 \frac{(E_{\alpha_0} - E_{\beta_0})^2\Gamma + 12\Gamma^3}{\pi (( E_{\alpha_0} - E_{\beta_0})^2 + 4\Gamma^2)^2} \leq \frac{3\omega_0 }{4\pi\Gamma}.
\end{split}
\end{equation}
Now, applying the sparsity condition, such that $\sum_{\alpha\beta}O_{\alpha\beta} \approx \sum_{\alpha}\sum_{n \in N_O}O_{\alpha, \alpha + n}$, we thus have,
\begin{equation}
\begin{split}
\sum_{\alpha_0, \beta_0} | \psi_{\alpha_0}\psi^*_{\beta_0} O_{\alpha_0\beta_0} | & \approx \sum_{\alpha_0, n} | \psi_{\alpha_0}\psi^*_{\alpha_0 + n} O_{\alpha_0\alpha_0 + n} | \\ &
\leq \max_{\alpha_0\beta_0}(O_{\alpha_0\beta_0}) \sum_{\alpha_0, n} | \psi_{\alpha_0}\psi^*_{\alpha_0 + n}  |\\& 
\leq \max_{\alpha_0\beta_0}(O_{\alpha_0\beta_0}) \sum_n \Bigg(\left( \sum_{\alpha_0} | \psi_{\alpha_0}|^2\right)\\& \qquad \times \left(\sum_{\alpha_0}|\psi^*_{\alpha_0 + n}|^2\right)
\Bigg)^{\frac{1}{2}}\\& = \max_{\alpha_0\beta_0}(O_{\alpha_0\beta_0}) N_O,
\end{split}
\end{equation}
where we have used the Cauchy-Schwarz inequality in the penultimate step. Thus, finally, we see that $|A(t)|$ is bounded for all time by
\begin{equation}
|A(t)| \leq  \max_{\alpha_0\beta_0}(O_{\alpha_0\beta_0}) N_O\frac{3\omega_0 }{4\pi\Gamma},
\end{equation}
which is small in comparison to other terms in the time evolution, and can thus be ignored.
  
\section{Time Dependence in Longditudinal and Transverse Fields}\label{App:Obs_x_field}

In the final case analyzed in the main text (see Fig. (\ref{fig:Gen_FDT})), we have an initial state $|\uparrow\rangle_S$, in the Hamiltonian $H_S = B_{z}^{(S)}\sigma_z^{(1)} + B_{x}^{(S)}\sigma_x^{(1)}$, and thus
\begin{equation}\label{eq:up_state1}
|\uparrow\rangle_S = \psi_+ |\phi_+\rangle_S + \psi_- |\phi_-\rangle_S,
\end{equation}
with 
\begin{equation}\label{eq:SB_coeffs1}
\begin{split}
& \psi_+ = \frac{B_z^{(S)} + E}{\sqrt{(B_z^{(S)} + E)^2 + (B_x^{(S)})^2}} \\& \psi_- = \frac{B_x^{(S)}}{\sqrt{(B_z^{(S)} + E)^2 + (B_x^{(S)})^2}},
\end{split}
\end{equation}
and $E := \sqrt{(B_z^{(S)})^2 + (B_x^{(S)})^2}$.
To obtain the full time dependence of the state in the Hamiltonian $H = H_S + H_B + H_{SB}$, from Eq. \eqref{eq:O_t}, we require the time evolution in the non-interacting part $\langle O(t) \rangle_0$. This is easily obtained, and is equal to 
\begin{equation}\label{eq:O_0}
\langle O(t) \rangle_0 = \sum_{\alpha_0, \beta_0} \psi_{\alpha_0}\psi^*_{\beta_0} O_{\alpha_0\beta_0} e^{-i(E_{\alpha_0} - E_{\beta_0} )t}, 
\end{equation}
with $\{\alpha_0\} = \{+, -\}$, and thus
\begin{equation}
\begin{split}
\langle \phi_+|\sigma_z|\phi_+\rangle & = - \langle \phi_-|\sigma_z|\phi_-\rangle \\& = \frac{(B_z^{(S)} + E)^2 - (B_x^{(S)})^2 }{(B_z^{(S)} + E)^2 + (B_x^{(S)})^2 },
\end{split}
\end{equation}
and
\begin{equation}
\begin{split}
\langle \phi_+|\sigma_z|\phi_-\rangle & = \langle \phi_-|\sigma_z|\phi_+\rangle \\& = -2\frac{(B_z^{(S)} + E)B_x^{(S)} }{(B_z^{(S)} + E)^2 + (B_x^{(S)})^2 }.\\&
\end{split}
\end{equation}
Then, from Eq. \eqref{eq:O_0}, we see that
\begin{equation}\label{eq:O_0t_fin}
\begin{split}
\langle O(t) \rangle_0 & = \left( \frac{(B_z^{(S)} + E)^2 - (B_x^{(S)})^2 }{(B_z^{(S)} + E)^2 + (B_x^{(S)})^2 } \right)^2 \\& + 4\left( \frac{(B_z^{(S)} + E)(B_x^{(S)})  }{(B_z^{(S)} + E)^2 + (B_x^{(S)})^2 } \right)^2 \cos{(2Et)}. 
\end{split}
\end{equation}
An example of this case is shown in Fig. (\ref{fig:Time_Dep}).\\

\section{Proof of Decoupling of Initial State and Observable Coefficients}\label{App:Flucs}
Here we prove the `decoupling' process required in Eqs. \eqref{eq:DE_fluc} and \eqref{eq:flucs_gen}, which may essentially be summarized by the statement that in the calculation of time-averaged fluctuations the coefficients $c_\mu(\alpha)$ contributed by the initial state may be considered independently of those in the observable elements $|O_{\mu\nu}|^2_{\mu\neq\nu}$, such that in the most general form we may replace
\begin{equation}\label{eq:decoupling_gen}
\begin{split}
\langle c_\mu(\alpha)&c_\nu(\beta)c_\mu(\alpha^\prime)c_\nu(\alpha^\prime) |O_{\mu\nu}|^2 \rangle_V \to \\& \langle c_\mu(\alpha)c_\nu(\beta)c_\mu(\alpha^\prime)c_\nu(\alpha^\prime) \rangle_V \langle |O_{\mu\nu}|^2 \rangle_V 
\end{split}
\end{equation}
in Eq. \eqref{eq:flucs_gen}.

We begin by discussing the form of correlation functions within the theory developed in Ref. \cite{Nation2018}, and note that below we explicitly discuss the off-diagonal, $\mu\neq \nu$, case, relevant for the time-averaged fluctuations. Using the method described here, we can in principle calculate any arbitrary correlation function from successive derivatives of the generating function, Eq. (\ref{eq:gen_func_1}), as shown in Eq. (\ref{eq:Derivs}). 
We can see from the generating function (\ref{eq:gen_func_1}), arbitrary correlation functions can be expressed in terms of products of two- and four-point correlation functions. Two point correlation functions are given by $\langle c_\mu(\alpha) c_\nu(\beta)\rangle = \Lambda(\mu, \alpha)\delta_{\mu\nu}\delta_{\alpha\beta}$, which is the same as one would expect for coefficients behaving as Gaussian distributed random variables of width $\Lambda(\mu, \alpha)$. Now, the four-point correlation function, Eq. (\ref{eq:Corr_Func}), may be seen as the sum of a Gaussian contraction,
\begin{equation}
\begin{split}
\acontraction{\langle c_\mu(}{\alpha}{)c_\nu(\beta)c_\mu (}{\alpha^\prime} \bcontraction{\langle c_\mu(\alpha) c_\nu(}{\beta}{)c_\mu(\alpha^\prime)c_\nu(}{\beta^\prime} \langle c_\mu(\alpha) c_\nu(\beta) c_\mu(\alpha^\prime) c_\nu(\beta^\prime)\rangle_V & \Rightarrow \langle c_\mu^2(\alpha)\rangle_V\langle c_\nu^2(\beta)\rangle_V\delta_{\alpha\alpha^\prime}\delta_{\beta\beta^\prime} \\& = \Lambda(\mu, \alpha) \Lambda(\nu, \beta) \delta_{\alpha\alpha^\prime}\delta_{\beta\beta^\prime},
\end{split}
\end{equation}
and non-Gaussian, or `four-leg', contractions, of which there are two:
\begin{widetext}
\begin{subequations}\label{eq:4legab}
\begin{eqnarray}
 \acontraction{\langle c_\mu(}{\alpha}{)c_\nu(}{\beta} \acontraction[1.25ex]{\langle c_\mu(}{\alpha}{)c_\nu(}{\beta} \bcontraction{\langle c_\mu(\alpha) c_\nu(\beta) c_\mu(}{\alpha^\prime}{)c_\nu((}{\beta} \bcontraction[1.25ex]{\langle c_\mu(\alpha) c_\nu(\beta) c_\mu(}{\alpha^\prime}{)c_\nu((}{\beta}\langle c_\mu(\alpha) c_\nu(\beta) c_\mu(\alpha^\prime) c_\nu(\beta^\prime)\rangle_V \Rightarrow \frac{\Lambda(\mu, \alpha)\Lambda(\nu, \alpha)\Lambda(\mu, \alpha^\prime)\Lambda(\nu, \alpha^\prime)\delta_{\alpha\beta}\delta_{\alpha^\prime\beta^\prime}}{\Lambda^{(2)}(\mu,\nu)} \\
 \bcontraction{\langle c_\mu(}{\alpha}{)c_\nu(\beta)c_\mu (\alpha^\prime)c_\nu(}{\beta}  \bcontraction[1.25ex]{\langle c_\mu(}{\alpha}{)c_\nu(\beta)c_\mu (\alpha^\prime)c_\nu(}{\beta}\acontraction{\langle c_\mu(\alpha) c_\nu(}{\beta}{)c_\mu(}{\beta} \acontraction[1.25ex]{\langle c_\mu(\alpha) c_\nu(}{\beta}{)c_\mu(}{\beta}\langle c_\mu(\alpha) c_\nu(\beta) c_\mu(\alpha^\prime) c_\nu(\beta^\prime)\rangle_V \Rightarrow \frac{\Lambda(\mu, \alpha)\Lambda(\nu, \alpha)\Lambda(\mu, \alpha^\prime)\Lambda(\nu, \alpha^\prime)\delta_{\alpha\beta^\prime}\delta_{\alpha^\prime\beta}}{\Lambda^{(2)}(\mu,\nu)}.
\end{eqnarray}
\end{subequations}
\end{widetext}
We reserve the double line contractions for the four-leg case. We note that the four-leg contractions arise as a consequence of enforcing the orthogonality of eigenstates of the random matrix Hamiltonian, such that if the $c_\mu(\alpha)$ coefficients were Gaussian distributed random numbers, as is commonly assumed, one would simply be left with the Gaussian contraction term. We further note that two point correlation functions are only explicitly required for correlation functions of $4n + 2 \, | \, n \in \mathbb{N}_0$ coefficients, as they are included here in the Gaussian contractions of the four-point correlation function.

Now, we wish to analyze the long-time fluctuations of a given observable $O$, defined in Eq. \eqref{eq:flucs_def}. 
In general, the initial state may be expressed as a superposition in the non-interacting basis:
\begin{equation}
|\psi(0)\rangle = \sum_{\alpha}\psi_\alpha|\phi_{\alpha}\rangle.
\end{equation}
We thus have, assuming non degenerate energy levels $E_\mu$ and energy gaps $E_\mu - E_\nu$,
\begin{equation}
\begin{split}
\delta_O^2(\infty) =\sum_{\alpha\beta\alpha^\prime\beta^\prime}&\sum_{\substack{\mu\nu\\ \mu\neq\nu}} \psi_\alpha\psi_\beta\psi_{\alpha^\prime}\psi_{\beta^\prime}  \\ & \times c_\mu(\alpha)c_\nu(\beta)c_\mu(\alpha^\prime)c_\nu(\beta^\prime)|O_{\mu\nu}|^2.\\&
\end{split}
\end{equation}
Now, assuming self averaging, we write
\begin{widetext}
\begin{equation}\label{eq:delta_SA}
\begin{split}
 \delta_O^2(\infty) & = \sum_{\alpha\beta\alpha^\prime\beta^\prime}\sum_{\substack{\mu\nu\\ \mu\neq\nu}}\psi_\alpha\psi_\beta\psi_{\alpha^\prime}\psi_{\beta^\prime} \langle c_\mu(\alpha)c_\nu(\beta)c_\mu(\alpha^\prime)c_\nu(\beta^\prime)|O_{\mu\nu}|^2\rangle_V \\& 
= \sum_{\alpha\beta\alpha^\prime\beta^\prime}\sum_{\alpha_1 \beta_1 \alpha_1^\prime \beta_1^\prime}\sum_{\substack{\mu\nu\\ \mu\neq\nu}}\psi_\alpha\psi_\beta\psi_{\alpha^\prime}\psi_{\beta^\prime}O_{\alpha_1\beta_1}O_{\alpha_1^\prime\beta_1^\prime} \times \langle c_\mu(\alpha)c_\nu(\beta)c_\mu(\alpha^\prime)c_\nu(\beta^\prime)c_\mu(\alpha_1)c_\nu(\beta_1)c_\mu(\alpha_1^\prime)c_\nu(\beta_1^\prime)\rangle_V, 
\end{split}
\end{equation}
\end{widetext}
which, if the initial state $|\psi(0)\rangle$ is a single eigenstate of $H_0$, $|\phi_{\alpha}\rangle$, we obtain simply
\begin{equation}\label{eq:delta_eig}
\delta_O^2(\infty) = \sum_{\substack{\mu\nu\\ \mu\neq\nu}} \langle |c_\mu(\alpha)|^2|c_\nu(\alpha)|^2|O_{\mu\nu}|^2\rangle_V.
\end{equation}

In principle, for a generic initial state, we thus require the calculation of an arbitrary 8-point correlation function, as seen in Eq. (\ref{eq:delta_SA}). We can see this requires four-leg contractions of all possible indices. We will observe, however, that the sections of the correlation function arising from the initial state coefficients (no subscript) and observable coefficients (subscript 1), decouple, and we obtain
\begin{equation}\label{eq:delta_eig2}
\delta_O^2(\infty) = \sum_{\substack{\mu\nu\\ \mu\neq\nu}} \langle |c_\mu(\alpha)|^2|c_\nu(\alpha)|^2\rangle_V\langle|O_{\mu\nu}|^2\rangle_V,
\end{equation}
such that only correlation functions within the respective coefficient types are required. We note that for generic initial states and observables this occurs as a consequence of the sparsity assumption. In the remainder of this section we introduce a method of contractions for four-point correlation functions in order to show this decoupling.

Suppose one wishes to evaluate the sum of correlation functions of initial state and observable coefficients
\begin{widetext}
\begin{equation}\label{eq:example_corr}
\sum_{\alpha\beta\alpha_1\beta_1\alpha_1\alpha_1^\prime}\langle c_\mu(\alpha)c_\nu(\alpha)c_\mu(\beta)c_\nu(\beta)c_\mu(\alpha_1)c_\nu(\alpha_1^\prime)c_\mu(\beta_1)c_\nu(\beta_1^\prime)\rangle_V,
\end{equation}
\end{widetext}
which, as discussed above, is made up of four point correlation functions of Gaussian, and four-leg contractions. One can see that an arbitrary four-point correlation function is of the order $\mathcal{O}\left(\left(\frac{\omega{0}}{\Gamma}\right)^\lambda\right)$, where $\lambda = N_\Lambda - N_\Sigma$, with $N_\Lambda$ the number of $\Lambda$ factors in the numerator minus the number of factors in the denominator, and $N_{\Sigma}$ is the number of summations. In this sense we have each $\Lambda$ contributing a factor on the order  $\mathcal{O}\left(\frac{\omega_0}{\Gamma}\right)$, and each summation contributing on the order  $\mathcal{O}\left(\frac{\Gamma}{\omega_0}\right)$. 


One can easily see in Eq. (\ref{eq:example_corr}), that particular contractions, Gaussian or non-Gaussian, in general reduce the number of summations over the non-interacting indices $\alpha, \beta, \cdots$. However, due to the repeated coefficient on the initial state side, contractions may be defined that require fewer summation restrictions, and thus these contractions dominate to lowest order in $\frac{\omega_0}{\Gamma}$. For example
\begin{widetext}
\begin{equation}\label{eq:within4}
 \acontraction[1.25ex]{\sum_{\alpha\beta\alpha_1\beta_1\alpha_1\alpha_1} \langle c_\mu(}{\beta}{)c_\nu(}{\beta}  
 \acontraction{\sum_{\alpha\beta\alpha_1\beta_1\alpha_1\alpha_1^\prime}\langle c_\mu(}{\beta}{)c_\nu(}{\beta}  
 \acontraction{\sum_{\alpha\beta\alpha_1\beta_1\alpha_1\alpha_1^\prime} \langle c_\mu(\alpha)c_\nu(\alpha)c_\mu(}{\beta}{)c_\nu(}{\beta} \acontraction[1.25ex]{\sum_{\alpha\beta\alpha_1\beta_1\alpha_1\alpha_1^\prime} \langle c_\mu(\alpha)c_\nu(\alpha)c_\mu(}{\beta}{)c_\nu(}{\beta}  \acontraction{\sum_{\alpha\beta\alpha_1\beta_1\alpha_1\alpha_1^\prime} \langle c_\mu(\alpha)c_\nu(\alpha)c_\mu(\beta)c_\nu(\beta)\rangle_V \langle c_\mu(}{\beta_1}{)c_\nu(}{\alpha_1^\prime} \acontraction[1.25ex]{\sum_{\alpha\beta\alpha_1\beta_1\alpha_1\alpha_1^\prime} \langle c_\mu(\alpha)c_\nu(\alpha)c_\mu(\beta)c_\nu(\beta)\rangle_V \langle c_\mu(}{\beta_1}{)c_\nu(}{\alpha_1^\prime} 
  \acontraction{\sum_{\alpha\beta\alpha_1\beta_1\alpha_1\alpha_1^\prime} \langle c_\mu(\alpha)c_\nu(\alpha)c_\mu(\beta)c_\nu(\beta)\rangle_V \langle c_\mu(\alpha_1)c_\nu(\alpha_1^\prime)c_\mu(}{\beta_1}{)c_\nu(}{\beta_1^\prime}
   \acontraction[1.25ex]{\sum_{\alpha\beta\alpha_1\beta_1\alpha_1\alpha_1^\prime} \langle c_\mu(\alpha)c_\nu(\alpha)c_\mu(\beta)c_\nu(\beta)\rangle_V \langle c_\mu(\alpha_1)c_\nu(\alpha_1^\prime)c_\mu(}{\beta_1}{)c_\nu(}{\beta_1^\prime} 
   \sum_{\alpha\beta\alpha_1\beta_1\stkout{\alpha_1^\prime}\stkout{\beta_1^\prime}} \langle c_\mu(\alpha)c_\nu(\alpha)c_\mu(\beta)c_\nu(\beta)\rangle_V \langle c_\mu(\alpha_1)c_\nu(\alpha_1^\prime)c_\mu(\beta_1)c_\nu(\beta_1^\prime)\rangle_V ,
\end{equation}
\end{widetext}
shows a single four-leg contraction for correlation functions within coefficient types, and
\begin{widetext}
\begin{equation}\label{eq:between4}
 \acontraction[1.25ex]{\sum_{\alpha\beta\alpha_1\beta_1\alpha_1\alpha_1} \langle c_\mu(}{\beta}{)c_\nu(}{\beta}  
 \acontraction{\sum_{\alpha\beta\alpha_1\beta_1\alpha_1\alpha_1^\prime}\langle c_\mu(}{\beta}{)c_\nu(}{\beta}  
 \acontraction{\sum_{\alpha\beta\alpha_1\beta_1\alpha_1\alpha_1^\prime} \langle c_\mu(\alpha)c_\nu(\alpha)c_\mu(}{\beta}{)c_\nu(}{\beta} \acontraction[1.25ex]{\sum_{\alpha\beta\alpha_1\beta_1\alpha_1\alpha_1^\prime} \langle c_\mu(\alpha)c_\nu(\alpha)c_\mu(}{\beta}{)c_\nu(}{\beta}  \acontraction{\sum_{\alpha\beta\alpha_1\beta_1\alpha_1\alpha_1^\prime} \langle c_\mu(\alpha)c_\nu(\alpha)c_\mu(\beta)c_\nu(\beta)\rangle_V \langle c_\mu(}{\beta_1}{)c_\nu(}{\alpha_1^\prime} \acontraction[1.25ex]{\sum_{\alpha\beta\alpha_1\beta_1\alpha_1\alpha_1^\prime} \langle c_\mu(\alpha)c_\nu(\alpha)c_\mu(\beta)c_\nu(\beta)\rangle_V \langle c_\mu(}{\beta_1}{)c_\nu(}{\alpha_1^\prime} 
  \acontraction{\sum_{\alpha\beta\alpha_1\beta_1\alpha_1\alpha_1^\prime} \langle c_\mu(\alpha)c_\nu(\alpha)c_\mu(\beta)c_\nu(\beta)\rangle_V \langle c_\mu(\alpha_1)c_\nu(\alpha_1^\prime)c_\mu(}{\beta_1}{)c_\nu(}{\beta_1^\prime}
   \acontraction[1.25ex]{\sum_{\alpha\beta\alpha_1\beta_1\alpha_1\alpha_1^\prime} \langle c_\mu(\alpha)c_\nu(\alpha)c_\mu(\beta)c_\nu(\beta)\rangle_V \langle c_\mu(\alpha_1)c_\nu(\alpha_1^\prime)c_\mu(}{\beta_1}{)c_\nu(}{\beta_1^\prime} 
   \sum_{\alpha\stkout{\beta}\alpha_1\beta_1\stkout{\alpha_1^\prime}\stkout{\beta_1^\prime}} \langle c_\mu(\alpha)c_\nu(\alpha)c_\mu(\beta)c_\nu(\beta_1^\prime)\rangle_V \langle c_\mu(\alpha_1)c_\nu(\alpha_1^\prime)c_\mu(\beta_1)c_\nu(\beta)\rangle_V ,
\end{equation}
\end{widetext}
similarly shows an example with a single coefficient swapped between types. The strikethroughs show the summations that are restricted due to the contractions. Note that due to the repeated coefficients in Eq. (\ref{eq:within4}) a four-leg contraction may be defined with no required restriction on summations. 

Now, for the simple case of Eq. (\ref{eq:delta_eig}), we can see that the required correlation functions have repeated indices in the initial state coefficients, and thus only contractions within coefficient types contribute. 

In the general case of Eq. (\ref{eq:delta_SA}) we have no such repeated indices. We thus employ the sparsity condition, that $O_{\alpha\beta}$ is in general sparse, and has a well defined form in the non-interacting basis, which is generally either diagonal, or has non-zero values at some energy $E_n$ from the diagonal. This can be easily seen for local observables made up of Pauli matrices. We thus replace $\sum_{\alpha\beta}O_{\alpha\beta} \to \sum_{\alpha}\sum_{n\in N_O}O_{\alpha, \alpha + n}$. We thus have 
\begin{widetext}
\begin{equation}\label{eq:delta_1}
\begin{split}
\delta_O^2(\infty) & = \sum_{\alpha\beta\alpha^\prime\beta^\prime}\sum_{\alpha_1 \alpha_1^\prime }\sum_{\substack{\mu\nu\\ \mu\neq\nu}} \sum_n \psi_\alpha\psi_\beta\psi_{\alpha^\prime}\psi_{\beta^\prime}O_{\alpha_1, \alpha_1 + n}O_{\alpha_1^\prime, \alpha_1^\prime + n} \\& \times \langle c_\mu(\alpha)c_\nu(\beta)c_\mu(\alpha^\prime)c_\nu(\beta^\prime)c_\mu(\alpha_1)c_\nu(\alpha_1 + n)c_\mu(\alpha_1^\prime)c_\nu(\alpha_1^\prime + n)\rangle_V,
\end{split}
\end{equation}
\end{widetext}
and therefore observe that we now have repeated indices in the observable type. We then see that, once again, contractions between coefficient types may be ignored to leading order. Thus, we obtain
\begin{widetext}
\begin{equation}\label{eq:delta_app}
\delta_O^2(\infty) = \sum_{\alpha\beta\alpha^\prime\beta^\prime}\sum_{\substack{\mu\nu\\ \mu\neq\nu}}\psi_\alpha\psi_\beta\psi_{\alpha^\prime}\psi_{\beta^\prime} \langle c_\mu(\alpha)c_\nu(\beta)c_\mu(\alpha^\prime)c_\nu(\beta^\prime)\rangle_V\langle|O_{\mu\nu}|^2\rangle_V.
\end{equation}
\end{widetext}
\section{QC-FDT for Arbitrary Initial States and Non-Diagonal Observables}\label{App:Generic_QCFDT}

After the simplification obtained by the method of contractions above, we see that correlations between initial state and observable factors of the time averaged fluctuations only contribute up to $\mathcal{O}\left( \left(\frac{\omega_0}{\Gamma}\right)^2\right)$, and thus may be ignored. As such, in the calculation of time-averaged fluctuations, Eq. (\ref{eq:delta_SA}), we may make the replacement Eq. \eqref{eq:decoupling_gen}, leading to a general form given by Eq. (\ref{eq:delta_app}). 

Considering initially the simplest generalization, the case of arbitrary initial states, where observables are diagonal in the non-interacting basis, $O_{\alpha\beta} \propto \delta_{\alpha\beta}$, we have
\begin{widetext}
\begin{equation}\label{eq:delta_full}
\begin{split}
\delta_O^2(\infty) & = \sum_{\alpha\beta}\sum_{\substack{\mu\nu\\ \mu\neq\nu}} |\psi_\alpha|^2|\psi_\beta|^2 \bigg[  \Lambda(\mu, \alpha)\Lambda(\nu, \beta)  - 2\frac{\Lambda(\mu, \alpha)\Lambda(\nu, \alpha)\Lambda(\mu, \beta)\Lambda(\nu, \beta)}{\Lambda^{(2)}(\mu, \nu)}   \bigg]  \overline{[\Delta O^2_{\alpha\alpha}]}_{\overline{\mu}} \Lambda^{(2)}(\mu, \nu) \\ & 
\approx \sum_{\alpha\beta}|\psi_\alpha|^2|\psi_\beta|^2\Lambda^{(4)}(\alpha, \beta) \overline{[\Delta O^2_{\alpha\alpha}]}_{\overline{\alpha}} - 2\sum_{\alpha\beta} \overline{[\Delta O^2_{\alpha\alpha}]}_{\overline{\alpha}}|\psi_\alpha|^2|\psi_\beta|^2\left(\Lambda^{(2)}(\alpha, \beta)\right)^2
\end{split}
\end{equation}
\end{widetext}
where $\overline{\alpha} = (\alpha+\beta)/2$, and we have used for the off-diagonal elements of $O$\cite{Nation2018},
\begin{equation}\label{eq:O_munu_0}
|O_{\mu,\nu}|^2_{\mu\neq\nu} = \overline{[\Delta O^2_{\alpha\alpha}]}_{\overline{\mu}} \Lambda^{(2)}(\mu, \nu).
\end{equation}
The summations over $\mu, \nu$ have been performed in Eq. (\ref{eq:delta_full}) by the prescription $\sum_{\mu}\to \int dE_\mu / \omega_0$, and the effective microcanonical average $ \overline{[\Delta O^2_{\alpha\alpha}]}_{\overline{\alpha}}$ is taken at the energy $(E_\alpha + E_\beta)/2$ via the smoothness property. Now, we may bound the second term by (using $\max(\Lambda^{(2)}(\alpha, \beta)) = \frac{\omega}{2\pi\Gamma}$), obtaining,
\begin{equation}
\begin{split}
2\sum_{\alpha\beta} & \overline{[\Delta O^2_{\alpha\alpha}]}_{\overline{\alpha}}|\psi_\alpha|^2|\psi_\beta|^2\left(\Lambda^{(2)}(\alpha, \beta)\right)^2 \\& \leq  2\sum_{\alpha\beta} \overline{[\Delta O^2_{\alpha\alpha}]}_{\overline{\alpha}} |\psi_\alpha|^2|\psi_\beta|^2 \frac{\omega_0^2}{4\pi^2\Gamma^2} \\& 
\leq \max_{\alpha_0 }\left(\overline{[\Delta O^2_{\alpha\alpha}]}_{{\alpha_0}}\right)\frac{\omega_0^2}{4\pi^2\Gamma^2},
 \end{split}
\end{equation}
which is on the order of $\omega_0^2$, and thus is negligible. 
Now, we have for arbitrary initial states,
\begin{equation}\label{eq:delta_arb_in1}
\delta_O^2(\infty) \approx  \overline{[\Delta O^2_{\alpha\alpha}]}_{\overline{\alpha_0}}\sum_{\alpha\beta}|\psi_\alpha|^2|\psi_\beta|^2\Lambda^{(4)}(\alpha, \beta).
\end{equation}
We note here that while this form of the QC-FDT looks rather different, one expects many typical initial states to show a very similar relation to the simpler form of $\delta^2 \sim \frac{1}{D(E)\Gamma}$. To illustrate this, we evaluate the relation (\ref{eq:delta_arb_in1}) for some example initial state distributions $\{\psi_\alpha\}$.

The first example we analyze is the case where $H_0$ itself may be split into interacting and non-interacting parts $H_0 = H_0^{(0)} + H_0^{(I)}$, where $H_0^{(I)}$ may be treated as a random matrix. In this case, the distribution of $|\psi_\alpha|^2$ is given by a Lorentzian of width $\Gamma_0$, $\Lambda_0(\overline{\mu}, \alpha)$, and thus
\begin{equation}
\begin{split}
\delta_O^2(\infty) & = \overline{[\Delta O^2_{\alpha\alpha}]}_{\overline{\alpha_0}}\frac{\omega_0 (4\Gamma + 2\Gamma_0) / \pi}{(4\Gamma + 2\Gamma_0)^2 } \\& = \overline{[\Delta O^2_{\alpha\alpha}]}_{\overline{\alpha_0}} \frac{\omega_0}{\pi(4\Gamma + 2\Gamma_0) },
\end{split}
\end{equation}
and we thus recover the CQ-FDT in the same form as for an initial state $|\phi_\alpha\rangle$, but with an altered effective width.

Next, we consider a bimodal distribution $|\psi(0)\rangle = \psi_\alpha|\phi_{\alpha_1}\rangle + \psi_\beta|\phi_{\alpha_2}\rangle$. Here we have
\begin{equation}
\delta_O^2(\infty) = \overline{[\Delta O^2_{\alpha\alpha}]}_{\overline{\alpha_0}}\frac{1}{2}\left( \frac{\omega_0}{ 4\pi\Gamma} + \Lambda^{(4)}(\alpha_1, \alpha_2)\right),
\end{equation}
which we can see resembles the simple case in the first term, and follows a Lorentzian distribution in the second. This reduces to the simple case for $E_{\alpha_1} - E_{\alpha_2} \ll \Gamma$. Continuing in the same manner, we see that we can rewrite the QC-FDT for an arbitrary distribution, $|\psi(0)\rangle = \sum_{\alpha} \psi_\alpha |\phi_\alpha\rangle$, as
\begin{equation}
\begin{split}
\delta_O^2(\infty) = & \overline{[\Delta O^2_{\alpha\alpha}]}_{\overline{\alpha_0}}\bigg( \sum_{\alpha} |\psi_\alpha|^4 \frac{\omega_0}{ 4\pi\Gamma} \\& + 2\sum_{\substack{\alpha\beta \\ \alpha>\beta}}|\psi_\alpha|^2|\psi_\beta|^2 \Lambda^{(4)}(\alpha, \beta)\bigg),
\end{split}
\end{equation} 
and thus we see that the contribution of the first term reduces substantially. Finally, for a microcanonical distribution $\psi_\alpha = 1/\sqrt{N^*} \, \forall \, \alpha \in [E_0 - \delta/2, E_0 + \delta/2]$, we have
\begin{equation}
\delta_O^2(\infty) = \overline{[\Delta O^2_{\alpha\alpha}]}_{\overline{\alpha_0}}\frac{1}{N^*}
\end{equation}
as $N^* \approx D(E_0) \delta$, ans assuming that $D(E)$ does not change much over the width $\delta$, we once again recover the QC-FDT in its original form. 

We may also analyze Eq. \eqref{eq:delta_arb_in1} using another example of the smoothness relation Eq. \eqref{smoothness.assumption}. We see that the summation
\begin{equation}
\sum_\alpha |\psi_\alpha|^2\Lambda^{(4)}(\alpha, \beta),
\end{equation}
may be obtained when the width of the distribution $\{|\psi_\alpha|^2\}$ is $\ll \Gamma$, we have that $\Lambda^{(4)}(\alpha, \beta)$ is essentially constant in this summation, such that $\sum_\alpha |\psi_\alpha|^2\Lambda^{(4)}(\alpha, \beta) \approx \Lambda^{(4)}(\alpha_0, \beta)\sum_\alpha |\psi_\alpha|^2 = \Lambda^{(4)}(\alpha_0, \beta)$. Repeating the same step with the sum over $\beta$, we obtain
\begin{equation}
\begin{split}
\delta_O^2(\infty) & \approx  \overline{[\Delta O^2_{\alpha\alpha}]}_{\overline{\alpha_0}}\Lambda^{(4)}(\alpha_0, \alpha_0)  \\&
= \overline{[\Delta O^2_{\alpha\alpha}]}_{\overline{\alpha_0}}\frac{\omega_0}{4\pi\Gamma},
\end{split}
\end{equation}
and thus the original QC-FDT is recovered.

We have thus observed that for many physical initial states we expect that Eq. (\ref{eq:delta_arb_in1}) reduces to the simpler form of $\delta \sim \frac{1}{D(E)\Gamma}$. 

We now turn our attention to observables that are not necessarily diagonal in the non-interacting basis, but fulfill instead the sparsity condition. Such observables were shown above to fulfill the ETH, and may be described by Eq. \eqref{eq:O_general}. Using this, as well as Eqs. (\ref{eq:delta_app}) and \eqref{eq:Corr_Func}, we have 
\begin{widetext}
\begin{equation}
\begin{split}
\delta_O^2(\infty) = & \sum_{\alpha\beta} |\psi_\alpha|^2|\psi_\beta|^2 \sum_{\substack{\mu\nu\\ \mu \neq \nu}} \Lambda(\mu, \alpha) \Lambda(\nu, \beta) \sum_n a_n \Lambda^{(2)}(\mu, \nu - n) \\ &
- 2\sum_{\alpha\beta} |\psi_\alpha|^2|\psi_\beta|^2 \sum_{\substack{\mu\nu\\ \mu \neq \nu}} \frac{\Lambda(\mu, \alpha) \Lambda(\nu, \alpha)\Lambda(\mu, \beta) \Lambda(\nu, \beta)}{\Lambda^{(2)}(\mu, \nu)} \sum_n a_n \Lambda^{(2)}(\mu, \nu - n),
\end{split}
\end{equation}
\end{widetext}
where the sum over $n$ is understood to be over the set ${\cal N}$. The second term can be seen to be bounded by
\begin{equation}
\begin{split}
2\sum_{\alpha\beta} |\psi_\alpha|^2|\psi_\beta|^2 \sum_{\substack{\mu\nu\\ \mu \neq \nu}} & \frac{\Lambda(\mu, \alpha) \Lambda(\nu, \alpha)\Lambda(\mu, \beta) \Lambda(\nu, \beta)}{\Lambda^{(2)}(\mu, \nu)} \\& \times \sum_n a_n \frac{\omega_0}{2\pi\Gamma},
\end{split}
\end{equation}
which in turn, using Eq. (\ref{eq:bound1}), and assuming $a_n(E_{\overline{\mu}})$ is essentially independent of $E_{\overline{\mu}}$ over a width $\Gamma$, is bounded by
\begin{equation}
\sum_n a_n \frac{3\omega_0^2}{4\pi\Gamma^2},
\end{equation}
and may thus be ignored. Now, as $\sum_{\mu}\Lambda(\mu, \alpha)\Lambda^{(2)}(\mu, \nu - n) = \Lambda^{(3)}(\nu - n, \alpha) = \Lambda^{(3)}(\nu, \alpha + n)$, and, similarly, $\sum_{\mu}\Lambda(\mu, \alpha + n)\Lambda^{(3)}(\mu, \beta) = \Lambda^{(4)}(\alpha, \beta - n)$, we have
\begin{equation}
\delta_O^2(\infty) = \sum_{\alpha\beta, n}a_n|\psi_\alpha|^2|\psi_\beta|^2\Lambda^{(4)}(\alpha, \beta - n),
\end{equation}
where $a_n$ is taken at the initial state energy. We now have the most general form of the QC-FDT. We note here that in order that the factor $a_n$ may be treated as both independent of $\mu, \nu$, and evaluated finally at the initial state energy $E_{\overline{\alpha_0}}$, requires that both $a_n(E_{\overline{\mu}})$ is a smooth function, approximately invariant over the width $\Gamma$ around this energy.

\section{QC-FDT For $\sigma_z$ in $B_x$ and $B_z$ Fields}\label{App:QCFDT_BX}

For an observable that is diagonal in the basis of eigenstates of the non-interacting Hamiltonian we have observed that the QC-FDT takes a remarkably simple form, which generalizes (see Appendix \ref{App:Generic_QCFDT}) to a similar relationship when these conditions are relaxed. In this section, we explicitly calculate the generalized case for the spin-chain system analyzed in the main text, given by 
\begin{equation}
H_S = B_{z}^{(S)}\sigma_z^{(1)} + B_{x}^{(S)}\sigma_x^{(1)},
\end{equation}
such that we have for an initial state $|\uparrow\rangle_S$, we have 
\begin{equation}\label{eq:up_state}
|\uparrow\rangle_S = \psi_+ |\phi_+\rangle_S + \psi_- |\phi_-\rangle_S,
\end{equation}
with 
\begin{equation}\label{eq:SB_coeffs}
\begin{split}
& \psi_+ = \frac{B_z^{(S)} + E}{\sqrt{(B_z^{(S)} + E)^2 + (B_x^{(S)})^2}} \\& \psi_- = \frac{B_x^{(S)}}{\sqrt{(B_z^{(S)} + E)^2 + (B_x^{(S)})^2}},
\end{split}
\end{equation}
and $E := \sqrt{(B_z^{(S)})^2 + (B_x^{(S)})^2}$. The eigenenergies are $\pm E$. Now, we find for the matrix elements of the observable $\sigma_z$,
\begin{equation}\label{eq:OppOmm}
\begin{split}
{}_S\langle \phi_+|\sigma_z|\phi_+\rangle_S & = - {}_S\langle \phi_-|\sigma_z|\phi_-\rangle_S \\&
= \frac{(B_z^{(S)} + E)^2 - (B_x^{(S)})^2 }{(B_z^{(S)} + E)^2 + (B_x^{(S)})^2 },
\end{split}
\end{equation}
and
\begin{equation}
\begin{split}
{}_S\langle \phi_+|\sigma_z|\phi_-\rangle_S & = {}_S\langle \phi_-|\sigma_z|\phi_+\rangle_S \\&
= - 2\frac{(B_z^{(S)} + E)B_x^{(S)} }{(B_z^{(S)} + E)^2 + (B_x^{(S)})^2 }.
\end{split}
\end{equation}
The relative value of the observable matrix elements dictates the relative height of the broadened peaks of the observable in the interacting basis $(|\sigma_z|^2)_{\mu\nu}$. The observable in the interacting basis is then, from Eq. (\ref{eq:O_general}), given by
\begin{equation}\label{eq:O_three_peaks}
\begin{split}
|O_{\mu\nu}|^2_{\mu\neq\nu} & = a_0 \Lambda^{(2)}(\mu, \nu) + a_1\Lambda^{(2)}(\mu, \nu + 2E) \\ & + a_2\Lambda^{(2)}(\mu, \nu - 2E) ,
\end{split}
\end{equation}
where $\{a_i\}_{i=0, 1, 2}$ are the respective height of the three peaks at energies $0, \pm 2E$. Thus, we have
\begin{equation}\label{eq:a0}
\begin{split}
a_0 & = \overline{[\Delta O_{\alpha\alpha}]}_{\overline{\alpha_0}} \\&
= \sum_{\alpha} \Lambda(\overline{\alpha_0}, \alpha)O^2_{\alpha\alpha} - \left( \sum_{\alpha}\Lambda(\overline{\alpha_0}, \alpha)O_{\alpha\alpha} \right)^2,
\end{split}
\end{equation}
where $\overline{[\Delta O_{\alpha\alpha}]}_{\overline{\alpha_0}}$ is evaluated at $\overline{\alpha_0}$ as it is the elements $O_{\mu \nu}$ around this energy that contribute to $\delta^2_O(\infty)$ in Eq. (\ref{eq:FDT_General}). Further, we note that the second term in Eq. (\ref{eq:a0}) can be identified with the square of the long-time average value of the observable, see Eq. (\ref{eq:O_t}). To evaluate the first term, we must understand the sum over $\alpha$ to also run over the bath states, in the sense that we may write
\begin{equation}
\sum_{\alpha}O_{\alpha\alpha} = \sum_{\alpha_+}O_{\alpha_+\alpha_+} + \sum_{\alpha_-}O_{\alpha_-\alpha_-},
\end{equation}
where $O_{\alpha_\pm\alpha_\pm} = {}_B\langle \phi_\alpha|{}_S\langle\phi_\pm|O|\phi_\pm\rangle_S|\phi_\alpha\rangle_B$. Using that $O = \sigma_z^{(S)} \otimes \mathbb{1}^{(B)}$, we have that $O_{\alpha\alpha} = {}_S\langle\phi_\pm|O|\phi_\pm\rangle_S$ does not explicitly depend on the bath state, and thus
\begin{equation}
\begin{split}
a_0 & = \sum_{\alpha_+}\Lambda(\overline{\alpha_0}, \alpha_+)|{}_S\langle \phi_+|\sigma_z|\phi_+\rangle_S|^2 \\& 
+ \sum_{\alpha_-}\Lambda(\overline{\alpha_0}, \alpha_-)|{}_S\langle \phi_-|\sigma_z|\phi_-\rangle_S|^2 - \left(\overline{\langle O(t)\rangle}\right)^2.
\end{split}
\end{equation}
Note that the bath degrees of freedom have an associated density of states that is half that of the whole system plus bath. Thus, we have $\sum_{\alpha_\pm}\Lambda(\overline{\alpha_0}, \alpha_\pm) \to \int \frac{dE}{\omega_0} \frac{\omega_02\Gamma / \pi}{(E_{\alpha_0} - E)^2 + (\Gamma)^2} = \frac{1}{2}$. Using Eq. (\ref{eq:OppOmm}), we thus have,
\begin{equation}
\begin{split}
a_0 = &\frac{1}{2}\Big(|{}_S\langle \phi_+|\sigma_z|\phi_+\rangle_S|^2 + |{}_S\langle \phi_-|\sigma_z|\phi_-\rangle_S|^2\Big) - \Big(\overline{\langle O(t)\rangle}\Big)^2 \\& = \frac{(B_z^{(S)})^2}{(B_z^{(S)})^2 + (B_x^{(S)})^2} - \left(\overline{\langle O(t)\rangle}\right)^2.
\end{split}
\end{equation}
A similar argument reveals,
\begin{equation}
a_1 = a_2 =\frac{1}{2} |{}_S\langle \phi_+|\sigma_z|\phi_-\rangle_S|^2 = \frac{1}{2}\frac{(B_x^{(S)})^2}{(B_z^{(S)})^2 + (B_x^{(S)})^2}.
\end{equation}
We note that this satisfies the sum rule $\sum_{\nu}|O_{\mu\nu}|^2 = (O^2)_{\mu\mu} = 1$, as $\sum_n a_n = \sum_{\nu \neq \mu}|O_{\mu\nu}|^2 = 1 - O_{\mu\mu}^2$, noting $O_{\mu\mu} = \sum_{\alpha\beta}c_\mu(\alpha)c_\mu(\beta)O_{\alpha\beta} = \sum_{\alpha}\Lambda(\mu, \alpha) O_{\alpha\alpha}$. Now, using Eqs. \eqref{eq:FDT_General}, (\ref{eq:up_state}), and (\ref{eq:O_three_peaks}), we obtain
\begin{widetext}
\begin{equation}
\begin{split}
\delta_{\sigma_z}^2(\infty) &= \frac{1}{D(E_{\overline{\alpha_0}})}\Bigg( \bigg(|\psi_+|^4 + |\psi_-|^4\bigg) \bigg(\frac{a_0}{4\pi\Gamma} + 2a_1\frac{4\Gamma/\pi}{(2E)^2 +(4\Gamma)^2}\bigg) \\&+ 2|\psi_+|^2|\psi_-|^2\bigg( a_0\frac{4\Gamma/\pi}{(2E)^2 +(4\Gamma)^2} +  \frac{a_1}{4\pi\Gamma} +  a_1\frac{4\Gamma/\pi}{(4E)^2 +(4\Gamma)^2}\bigg)\Bigg).
\end{split}
\end{equation}
\end{widetext}
Here we note that in Eq. \eqref{eq:FDT_General} $\Lambda(\alpha, \beta + n)$ is a function of $E_\alpha - E_\beta + E_n$, with $E_\alpha - E_\beta$ giving the possible values $0, \pm 2E$. $E_n$ has the same possible values, as it labels the peak energies of the observable. Observe that in various physical limits we also recover the QC-FDT of the simpler form $\delta \sim \frac{1}{\Gamma}$, for example, when $E \gg \Gamma$, the Lorentzian terms are small, and the original scaling is obtained. In fact, as with the case for diagonal observables and general initial states, we expect this simpler form to hold up to a factor for most cases.

We further comment that in the generalized case the assumption that the density of states does not change over the relevant widths is not always valid, and may cause deviations from the result above by the effective rescaling of the $a_n$ factors for large $E_n$. This occurs as the implicit assumption is now $\Gamma < E_n < W$, where $W$ is the characteristic width of the density of states.
\bibliographystyle{apsrev4-1}
\bibliography{bibli}

\begin{thebibliography}{43}%
\makeatletter
\providecommand \@ifxundefined [1]{%
 \@ifx{#1\undefined}
}%
\providecommand \@ifnum [1]{%
 \ifnum #1\expandafter \@firstoftwo
 \else \expandafter \@secondoftwo
 \fi
}%
\providecommand \@ifx [1]{%
 \ifx #1\expandafter \@firstoftwo
 \else \expandafter \@secondoftwo
 \fi
}%
\providecommand \natexlab [1]{#1}%
\providecommand \enquote  [1]{``#1''}%
\providecommand \bibnamefont  [1]{#1}%
\providecommand \bibfnamefont [1]{#1}%
\providecommand \citenamefont [1]{#1}%
\providecommand \href@noop [0]{\@secondoftwo}%
\providecommand \href [0]{\begingroup \@sanitize@url \@href}%
\providecommand \@href[1]{\@@startlink{#1}\@@href}%
\providecommand \@@href[1]{\endgroup#1\@@endlink}%
\providecommand \@sanitize@url [0]{\catcode `\\12\catcode `\$12\catcode
  `\&12\catcode `\#12\catcode `\^12\catcode `\_12\catcode `\%12\relax}%
\providecommand \@@startlink[1]{}%
\providecommand \@@endlink[0]{}%
\providecommand \url  [0]{\begingroup\@sanitize@url \@url }%
\providecommand \@url [1]{\endgroup\@href {#1}{\urlprefix }}%
\providecommand \urlprefix  [0]{URL }%
\providecommand \Eprint [0]{\href }%
\providecommand \doibase [0]{http://dx.doi.org/}%
\providecommand \selectlanguage [0]{\@gobble}%
\providecommand \bibinfo  [0]{\@secondoftwo}%
\providecommand \bibfield  [0]{\@secondoftwo}%
\providecommand \translation [1]{[#1]}%
\providecommand \BibitemOpen [0]{}%
\providecommand \bibitemStop [0]{}%
\providecommand \bibitemNoStop [0]{.\EOS\space}%
\providecommand \EOS [0]{\spacefactor3000\relax}%
\providecommand \BibitemShut  [1]{\csname bibitem#1\endcsname}%
\let\auto@bib@innerbib\@empty
\bibitem [{\citenamefont {Neumann}(2010)}]{Neumann2010}%
  \BibitemOpen
  \bibfield  {author} {\bibinfo {author} {\bibfnamefont {J.~V.}\ \bibnamefont
  {Neumann}},\ }\href {\doibase 10.1140/epjh/e2010-00008-5} {\bibfield
  {journal} {\bibinfo  {journal} {The European Physical Journal H}\ }\textbf
  {\bibinfo {volume} {237}},\ \bibinfo {pages} {41} (\bibinfo {year}
  {2010})}\BibitemShut {NoStop}%
\bibitem [{\citenamefont {Rigol}\ \emph {et~al.}(2008)\citenamefont {Rigol},
  \citenamefont {Dunjko},\ and\ \citenamefont {Olshanii}}]{Rigol2008}%
  \BibitemOpen
  \bibfield  {author} {\bibinfo {author} {\bibfnamefont {M.}~\bibnamefont
  {Rigol}}, \bibinfo {author} {\bibfnamefont {V.}~\bibnamefont {Dunjko}}, \
  and\ \bibinfo {author} {\bibfnamefont {M.}~\bibnamefont {Olshanii}},\ }\href
  {\doibase 10.1038/nature06838} {\bibfield  {journal} {\bibinfo  {journal}
  {Nature}\ }\textbf {\bibinfo {volume} {452}},\ \bibinfo {pages} {854}
  (\bibinfo {year} {2008})}\BibitemShut {NoStop}%
\bibitem [{\citenamefont {D'Alessio}\ \emph {et~al.}(2016)\citenamefont
  {D'Alessio}, \citenamefont {Kafri}, \citenamefont {Polkovnikov},\ and\
  \citenamefont {Rigol}}]{DAlessio2016}%
  \BibitemOpen
  \bibfield  {author} {\bibinfo {author} {\bibfnamefont {L.}~\bibnamefont
  {D'Alessio}}, \bibinfo {author} {\bibfnamefont {Y.}~\bibnamefont {Kafri}},
  \bibinfo {author} {\bibfnamefont {A.}~\bibnamefont {Polkovnikov}}, \ and\
  \bibinfo {author} {\bibfnamefont {M.}~\bibnamefont {Rigol}},\ }\href
  {\doibase 10.1080/00018732.2016.1198134} {\bibfield  {journal} {\bibinfo
  {journal} {Advances in Physics}\ }\textbf {\bibinfo {volume} {65}},\ \bibinfo
  {pages} {239} (\bibinfo {year} {2016})}\BibitemShut {NoStop}%
\bibitem [{\citenamefont {Gogolin}\ and\ \citenamefont
  {Eisert}(2016)}]{Gogolin2016}%
  \BibitemOpen
  \bibfield  {author} {\bibinfo {author} {\bibfnamefont {C.}~\bibnamefont
  {Gogolin}}\ and\ \bibinfo {author} {\bibfnamefont {J.}~\bibnamefont
  {Eisert}},\ }\href {\doibase 10.1088/0034-4885/79/5/056001} {\bibfield
  {journal} {\bibinfo  {journal} {Rep. Prog. Phys.}\ }\textbf {\bibinfo
  {volume} {79}},\ \bibinfo {pages} {056001} (\bibinfo {year}
  {2016})}\BibitemShut {NoStop}%
\bibitem [{\citenamefont {{Borgonovi}}\ \emph {et~al.}(2016)\citenamefont
  {{Borgonovi}}, \citenamefont {{Izrailev}}, \citenamefont {{Santos}},\ and\
  \citenamefont {{Zelevinsky}}}]{Borgonovi2016}%
  \BibitemOpen
  \bibfield  {author} {\bibinfo {author} {\bibfnamefont {F.}~\bibnamefont
  {{Borgonovi}}}, \bibinfo {author} {\bibfnamefont {F.~M.}\ \bibnamefont
  {{Izrailev}}}, \bibinfo {author} {\bibfnamefont {L.~F.}\ \bibnamefont
  {{Santos}}}, \ and\ \bibinfo {author} {\bibfnamefont {V.~G.}\ \bibnamefont
  {{Zelevinsky}}},\ }\href {\doibase 10.1016/j.physrep.2016.02.005} {\bibfield
  {journal} {\bibinfo  {journal} {Physics Reports}\ }\textbf {\bibinfo {volume}
  {626}},\ \bibinfo {pages} {1} (\bibinfo {year} {2016})}\BibitemShut {NoStop}%
\bibitem [{\citenamefont {{Mori}}\ \emph {et~al.}(2018)\citenamefont {{Mori}},
  \citenamefont {{Ikeda}}, \citenamefont {{Kaminishi}},\ and\ \citenamefont
  {{Ueda}}}]{Mori2018}%
  \BibitemOpen
  \bibfield  {author} {\bibinfo {author} {\bibfnamefont {T.}~\bibnamefont
  {{Mori}}}, \bibinfo {author} {\bibfnamefont {T.~N.}\ \bibnamefont {{Ikeda}}},
  \bibinfo {author} {\bibfnamefont {E.}~\bibnamefont {{Kaminishi}}}, \ and\
  \bibinfo {author} {\bibfnamefont {M.}~\bibnamefont {{Ueda}}},\ }\href
  {\doibase 10.1088/1361-6455/aabcdf} {\bibfield  {journal} {\bibinfo
  {journal} {J. Phys. B: Atomic Molecular Physics}\ }\textbf {\bibinfo {volume}
  {51}},\ \bibinfo {eid} {112001} (\bibinfo {year} {2018})}\BibitemShut
  {NoStop}%
\bibitem [{\citenamefont {Schreiber}\ \emph {et~al.}(2015)\citenamefont
  {Schreiber}, \citenamefont {Hodgman}, \citenamefont {Bordia}, \citenamefont
  {L{\"{u}}schen}, \citenamefont {Fischer}, \citenamefont {Vosk}, \citenamefont
  {Altman}, \citenamefont {Schneider},\ and\ \citenamefont
  {Bloch}}]{Schreiber2015}%
  \BibitemOpen
  \bibfield  {author} {\bibinfo {author} {\bibfnamefont {M.}~\bibnamefont
  {Schreiber}}, \bibinfo {author} {\bibfnamefont {S.~S.}\ \bibnamefont
  {Hodgman}}, \bibinfo {author} {\bibfnamefont {P.}~\bibnamefont {Bordia}},
  \bibinfo {author} {\bibfnamefont {H.~P.}\ \bibnamefont {L{\"{u}}schen}},
  \bibinfo {author} {\bibfnamefont {M.~H.}\ \bibnamefont {Fischer}}, \bibinfo
  {author} {\bibfnamefont {R.}~\bibnamefont {Vosk}}, \bibinfo {author}
  {\bibfnamefont {E.}~\bibnamefont {Altman}}, \bibinfo {author} {\bibfnamefont
  {U.}~\bibnamefont {Schneider}}, \ and\ \bibinfo {author} {\bibfnamefont
  {I.}~\bibnamefont {Bloch}},\ }\href {\doibase 10.1126/science.aaa7432}
  {\bibfield  {journal} {\bibinfo  {journal} {Science}\ }\textbf {\bibinfo
  {volume} {349}},\ \bibinfo {pages} {842} (\bibinfo {year}
  {2015})}\BibitemShut {NoStop}%
\bibitem [{\citenamefont {Clos}\ \emph {et~al.}(2016)\citenamefont {Clos},
  \citenamefont {Porras}, \citenamefont {Warring},\ and\ \citenamefont
  {Schaetz}}]{Clos2016a}%
  \BibitemOpen
  \bibfield  {author} {\bibinfo {author} {\bibfnamefont {G.}~\bibnamefont
  {Clos}}, \bibinfo {author} {\bibfnamefont {D.}~\bibnamefont {Porras}},
  \bibinfo {author} {\bibfnamefont {U.}~\bibnamefont {Warring}}, \ and\
  \bibinfo {author} {\bibfnamefont {T.}~\bibnamefont {Schaetz}},\ }\href
  {\doibase 10.1103/PhysRevLett.117.170401} {\bibfield  {journal} {\bibinfo
  {journal} {Phys. Rev. Lett.}\ }\textbf {\bibinfo {volume} {117}},\ \bibinfo
  {pages} {170401} (\bibinfo {year} {2016})}\BibitemShut {NoStop}%
\bibitem [{\citenamefont {Kaufman}\ \emph {et~al.}(2016)\citenamefont
  {Kaufman}, \citenamefont {Tai}, \citenamefont {Lukin}, \citenamefont
  {Rispoli}, \citenamefont {Schittko}, \citenamefont {Preiss},\ and\
  \citenamefont {Greiner}}]{Kaufman2016}%
  \BibitemOpen
  \bibfield  {author} {\bibinfo {author} {\bibfnamefont {A.~M.}\ \bibnamefont
  {Kaufman}}, \bibinfo {author} {\bibfnamefont {M.~E.}\ \bibnamefont {Tai}},
  \bibinfo {author} {\bibfnamefont {A.}~\bibnamefont {Lukin}}, \bibinfo
  {author} {\bibfnamefont {M.}~\bibnamefont {Rispoli}}, \bibinfo {author}
  {\bibfnamefont {R.}~\bibnamefont {Schittko}}, \bibinfo {author}
  {\bibfnamefont {P.~M.}\ \bibnamefont {Preiss}}, \ and\ \bibinfo {author}
  {\bibfnamefont {M.}~\bibnamefont {Greiner}},\ }\href {\doibase
  10.1126/science.aaf7894} {\bibfield  {journal} {\bibinfo  {journal}
  {Science}\ }\textbf {\bibinfo {volume} {353}},\ \bibinfo {pages} {794}
  (\bibinfo {year} {2016})}\BibitemShut {NoStop}%
\bibitem [{\citenamefont {Neill}\ \emph {et~al.}(2016)\citenamefont {Neill},
  \citenamefont {Roushan}, \citenamefont {Fang}, \citenamefont {Chen},
  \citenamefont {Kolodrubetz}, \citenamefont {Chen}, \citenamefont {Megrant},
  \citenamefont {Barends}, \citenamefont {Campbell}, \citenamefont {Chiaro},
  \citenamefont {Dunsworth}, \citenamefont {Jeffrey}, \citenamefont {Kelly},
  \citenamefont {Mutus}, \citenamefont {O'Malley}, \citenamefont {Quintana},
  \citenamefont {Sank}, \citenamefont {Vainsencher}, \citenamefont {Wenner},
  \citenamefont {White}, \citenamefont {Polkovnikov},\ and\ \citenamefont
  {Martinis}}]{Neill2016}%
  \BibitemOpen
  \bibfield  {author} {\bibinfo {author} {\bibfnamefont {C.}~\bibnamefont
  {Neill}}, \bibinfo {author} {\bibfnamefont {P.}~\bibnamefont {Roushan}},
  \bibinfo {author} {\bibfnamefont {M.}~\bibnamefont {Fang}}, \bibinfo {author}
  {\bibfnamefont {Y.}~\bibnamefont {Chen}}, \bibinfo {author} {\bibfnamefont
  {M.}~\bibnamefont {Kolodrubetz}}, \bibinfo {author} {\bibfnamefont
  {Z.}~\bibnamefont {Chen}}, \bibinfo {author} {\bibfnamefont {A.}~\bibnamefont
  {Megrant}}, \bibinfo {author} {\bibfnamefont {R.}~\bibnamefont {Barends}},
  \bibinfo {author} {\bibfnamefont {B.}~\bibnamefont {Campbell}}, \bibinfo
  {author} {\bibfnamefont {B.}~\bibnamefont {Chiaro}}, \bibinfo {author}
  {\bibfnamefont {A.}~\bibnamefont {Dunsworth}}, \bibinfo {author}
  {\bibfnamefont {E.}~\bibnamefont {Jeffrey}}, \bibinfo {author} {\bibfnamefont
  {J.}~\bibnamefont {Kelly}}, \bibinfo {author} {\bibfnamefont
  {J.}~\bibnamefont {Mutus}}, \bibinfo {author} {\bibfnamefont {P.~J.}\
  \bibnamefont {O'Malley}}, \bibinfo {author} {\bibfnamefont {C.}~\bibnamefont
  {Quintana}}, \bibinfo {author} {\bibfnamefont {D.}~\bibnamefont {Sank}},
  \bibinfo {author} {\bibfnamefont {A.}~\bibnamefont {Vainsencher}}, \bibinfo
  {author} {\bibfnamefont {J.}~\bibnamefont {Wenner}}, \bibinfo {author}
  {\bibfnamefont {T.~C.}\ \bibnamefont {White}}, \bibinfo {author}
  {\bibfnamefont {A.}~\bibnamefont {Polkovnikov}}, \ and\ \bibinfo {author}
  {\bibfnamefont {J.~M.}\ \bibnamefont {Martinis}},\ }\href {\doibase
  10.1038/nphys3830} {\bibfield  {journal} {\bibinfo  {journal} {Nature
  Physics}\ }\textbf {\bibinfo {volume} {12}},\ \bibinfo {pages} {1037}
  (\bibinfo {year} {2016})}\BibitemShut {NoStop}%
\bibitem [{\citenamefont {Neill}\ \emph {et~al.}(2018)\citenamefont {Neill},
  \citenamefont {Roushan}, \citenamefont {Kechedzhi}, \citenamefont {Boixo},
  \citenamefont {Isakov}, \citenamefont {Smelyanskiy}, \citenamefont {Megrant},
  \citenamefont {Chiaro}, \citenamefont {Dunsworth}, \citenamefont {Arya},
  \citenamefont {Barends}, \citenamefont {Burkett}, \citenamefont {Chen},
  \citenamefont {Chen}, \citenamefont {Fowler}, \citenamefont {Foxen},
  \citenamefont {Giustina}, \citenamefont {Graff}, \citenamefont {Jeffrey},
  \citenamefont {Huang}, \citenamefont {Kelly}, \citenamefont {Klimov},
  \citenamefont {Lucero}, \citenamefont {Mutus}, \citenamefont {Neeley},
  \citenamefont {Quintana}, \citenamefont {Sank}, \citenamefont {Vainsencher},
  \citenamefont {Wenner}, \citenamefont {White}, \citenamefont {Neven},\ and\
  \citenamefont {Martinis}}]{Neill2018}%
  \BibitemOpen
  \bibfield  {author} {\bibinfo {author} {\bibfnamefont {C.}~\bibnamefont
  {Neill}}, \bibinfo {author} {\bibfnamefont {P.}~\bibnamefont {Roushan}},
  \bibinfo {author} {\bibfnamefont {K.}~\bibnamefont {Kechedzhi}}, \bibinfo
  {author} {\bibfnamefont {S.}~\bibnamefont {Boixo}}, \bibinfo {author}
  {\bibfnamefont {S.~V.}\ \bibnamefont {Isakov}}, \bibinfo {author}
  {\bibfnamefont {V.}~\bibnamefont {Smelyanskiy}}, \bibinfo {author}
  {\bibfnamefont {A.}~\bibnamefont {Megrant}}, \bibinfo {author} {\bibfnamefont
  {B.}~\bibnamefont {Chiaro}}, \bibinfo {author} {\bibfnamefont
  {A.}~\bibnamefont {Dunsworth}}, \bibinfo {author} {\bibfnamefont
  {K.}~\bibnamefont {Arya}}, \bibinfo {author} {\bibfnamefont {R.}~\bibnamefont
  {Barends}}, \bibinfo {author} {\bibfnamefont {B.}~\bibnamefont {Burkett}},
  \bibinfo {author} {\bibfnamefont {Y.}~\bibnamefont {Chen}}, \bibinfo {author}
  {\bibfnamefont {Z.}~\bibnamefont {Chen}}, \bibinfo {author} {\bibfnamefont
  {A.}~\bibnamefont {Fowler}}, \bibinfo {author} {\bibfnamefont
  {B.}~\bibnamefont {Foxen}}, \bibinfo {author} {\bibfnamefont
  {M.}~\bibnamefont {Giustina}}, \bibinfo {author} {\bibfnamefont
  {R.}~\bibnamefont {Graff}}, \bibinfo {author} {\bibfnamefont
  {E.}~\bibnamefont {Jeffrey}}, \bibinfo {author} {\bibfnamefont
  {T.}~\bibnamefont {Huang}}, \bibinfo {author} {\bibfnamefont
  {J.}~\bibnamefont {Kelly}}, \bibinfo {author} {\bibfnamefont
  {P.}~\bibnamefont {Klimov}}, \bibinfo {author} {\bibfnamefont
  {E.}~\bibnamefont {Lucero}}, \bibinfo {author} {\bibfnamefont
  {J.}~\bibnamefont {Mutus}}, \bibinfo {author} {\bibfnamefont
  {M.}~\bibnamefont {Neeley}}, \bibinfo {author} {\bibfnamefont
  {C.}~\bibnamefont {Quintana}}, \bibinfo {author} {\bibfnamefont
  {D.}~\bibnamefont {Sank}}, \bibinfo {author} {\bibfnamefont {A.}~\bibnamefont
  {Vainsencher}}, \bibinfo {author} {\bibfnamefont {J.}~\bibnamefont {Wenner}},
  \bibinfo {author} {\bibfnamefont {T.~C.}\ \bibnamefont {White}}, \bibinfo
  {author} {\bibfnamefont {H.}~\bibnamefont {Neven}}, \ and\ \bibinfo {author}
  {\bibfnamefont {J.~M.}\ \bibnamefont {Martinis}},\ }\href {\doibase
  10.1126/science.aao4309} {\bibfield  {journal} {\bibinfo  {journal}
  {Science}\ }\textbf {\bibinfo {volume} {360}},\ \bibinfo {pages} {195}
  (\bibinfo {year} {2018})}\BibitemShut {NoStop}%
\bibitem [{\citenamefont {Popescu}\ \emph {et~al.}(2006)\citenamefont
  {Popescu}, \citenamefont {Short},\ and\ \citenamefont
  {Winter}}]{Popescu2006}%
  \BibitemOpen
  \bibfield  {author} {\bibinfo {author} {\bibfnamefont {S.}~\bibnamefont
  {Popescu}}, \bibinfo {author} {\bibfnamefont {A.~J.}\ \bibnamefont {Short}},
  \ and\ \bibinfo {author} {\bibfnamefont {A.}~\bibnamefont {Winter}},\ }\href
  {\doibase 10.1038/nphys444} {\bibfield  {journal} {\bibinfo  {journal}
  {Nature Physics}\ }\textbf {\bibinfo {volume} {2}},\ \bibinfo {pages} {754}
  (\bibinfo {year} {2006})}\BibitemShut {NoStop}%
\bibitem [{\citenamefont {Reimann}(2007)}]{Reimann2008}%
  \BibitemOpen
  \bibfield  {author} {\bibinfo {author} {\bibfnamefont {P.}~\bibnamefont
  {Reimann}},\ }\href {\doibase 10.1103/PhysRevLett.99.160404} {\bibfield
  {journal} {\bibinfo  {journal} {Phys. Rev. Lett.}\ }\textbf {\bibinfo
  {volume} {99}},\ \bibinfo {pages} {160404} (\bibinfo {year}
  {2007})}\BibitemShut {NoStop}%
\bibitem [{\citenamefont {Linden}\ \emph {et~al.}(2009)\citenamefont {Linden},
  \citenamefont {Popescu}, \citenamefont {Short},\ and\ \citenamefont
  {Winter}}]{Linden2009}%
  \BibitemOpen
  \bibfield  {author} {\bibinfo {author} {\bibfnamefont {N.}~\bibnamefont
  {Linden}}, \bibinfo {author} {\bibfnamefont {S.}~\bibnamefont {Popescu}},
  \bibinfo {author} {\bibfnamefont {A.~J.}\ \bibnamefont {Short}}, \ and\
  \bibinfo {author} {\bibfnamefont {A.}~\bibnamefont {Winter}},\ }\href
  {\doibase 10.1103/PhysRevE.79.061103} {\bibfield  {journal} {\bibinfo
  {journal} {Phys. Rev. E}\ }\textbf {\bibinfo {volume} {79}},\ \bibinfo
  {pages} {061103} (\bibinfo {year} {2009})}\BibitemShut {NoStop}%
\bibitem [{\citenamefont {Bartsch}\ and\ \citenamefont
  {Gemmer}(2009)}]{Bartsch2009}%
  \BibitemOpen
  \bibfield  {author} {\bibinfo {author} {\bibfnamefont {C.}~\bibnamefont
  {Bartsch}}\ and\ \bibinfo {author} {\bibfnamefont {J.}~\bibnamefont
  {Gemmer}},\ }\href {https://link.aps.org/doi/10.1103/PhysRevLett.102.110403}
  {\bibfield  {journal} {\bibinfo  {journal} {Phys. Rev. Lett.}\ }\textbf
  {\bibinfo {volume} {102}},\ \bibinfo {pages} {110403} (\bibinfo {year}
  {2009})}\BibitemShut {NoStop}%
\bibitem [{\citenamefont {Deutsch}(1991{\natexlab{a}})}]{Deutsch1991}%
  \BibitemOpen
  \bibfield  {author} {\bibinfo {author} {\bibfnamefont {J.~M.}\ \bibnamefont
  {Deutsch}},\ }\href {\doibase 10.1103/PhysRevA.43.2046} {\bibfield  {journal}
  {\bibinfo  {journal} {Physical Review A}\ }\textbf {\bibinfo {volume} {43}},\
  \bibinfo {pages} {2046} (\bibinfo {year} {1991}{\natexlab{a}})}\BibitemShut
  {NoStop}%
\bibitem [{\citenamefont {Srednicki}(1999)}]{Srednicki1999}%
  \BibitemOpen
  \bibfield  {author} {\bibinfo {author} {\bibfnamefont {M.}~\bibnamefont
  {Srednicki}},\ }\href {\doibase 10.1088/0305-4470/32/7/007} {\bibfield
  {journal} {\bibinfo  {journal} {J. Phys. A: Math. Gen.}\ }\textbf {\bibinfo
  {volume} {32}},\ \bibinfo {pages} {1163} (\bibinfo {year}
  {1999})}\BibitemShut {NoStop}%
\bibitem [{\citenamefont {Santos}\ and\ \citenamefont
  {Rigol}(2010{\natexlab{a}})}]{Santos2010}%
  \BibitemOpen
  \bibfield  {author} {\bibinfo {author} {\bibfnamefont {L.~F.}\ \bibnamefont
  {Santos}}\ and\ \bibinfo {author} {\bibfnamefont {M.}~\bibnamefont {Rigol}},\
  }\href {\doibase 10.1103/PhysRevE.81.036206} {\bibfield  {journal} {\bibinfo
  {journal} {Phys. Rev. E}\ }\textbf {\bibinfo {volume} {81}},\ \bibinfo
  {pages} {036206} (\bibinfo {year} {2010}{\natexlab{a}})}\BibitemShut
  {NoStop}%
\bibitem [{\citenamefont {Santos}\ and\ \citenamefont
  {Rigol}(2010{\natexlab{b}})}]{Santos2010a}%
  \BibitemOpen
  \bibfield  {author} {\bibinfo {author} {\bibfnamefont {L.~F.}\ \bibnamefont
  {Santos}}\ and\ \bibinfo {author} {\bibfnamefont {M.}~\bibnamefont {Rigol}},\
  }\href {\doibase 10.1103/PhysRevE.82.031130} {\bibfield  {journal} {\bibinfo
  {journal} {Phys. Rev. E}\ }\textbf {\bibinfo {volume} {82}},\ \bibinfo
  {pages} {031130} (\bibinfo {year} {2010}{\natexlab{b}})}\BibitemShut
  {NoStop}%
\bibitem [{\citenamefont {Steinigeweg}\ \emph {et~al.}(2014)\citenamefont
  {Steinigeweg}, \citenamefont {Khodja}, \citenamefont {Niemeyer},
  \citenamefont {Gogolin},\ and\ \citenamefont {Gemmer}}]{Steinigeweg2013}%
  \BibitemOpen
  \bibfield  {author} {\bibinfo {author} {\bibfnamefont {R.}~\bibnamefont
  {Steinigeweg}}, \bibinfo {author} {\bibfnamefont {A.}~\bibnamefont {Khodja}},
  \bibinfo {author} {\bibfnamefont {H.}~\bibnamefont {Niemeyer}}, \bibinfo
  {author} {\bibfnamefont {C.}~\bibnamefont {Gogolin}}, \ and\ \bibinfo
  {author} {\bibfnamefont {J.}~\bibnamefont {Gemmer}},\ }\href {\doibase
  10.1103/PhysRevLett.112.130403} {\bibfield  {journal} {\bibinfo  {journal}
  {Phys. Rev. Lett.}\ }\textbf {\bibinfo {volume} {112}},\ \bibinfo {pages}
  {130403} (\bibinfo {year} {2014})}\BibitemShut {NoStop}%
\bibitem [{\citenamefont {Beugeling}\ \emph {et~al.}(2014)\citenamefont
  {Beugeling}, \citenamefont {Moessner},\ and\ \citenamefont
  {Haque}}]{Beugeling2014}%
  \BibitemOpen
  \bibfield  {author} {\bibinfo {author} {\bibfnamefont {W.}~\bibnamefont
  {Beugeling}}, \bibinfo {author} {\bibfnamefont {R.}~\bibnamefont {Moessner}},
  \ and\ \bibinfo {author} {\bibfnamefont {M.}~\bibnamefont {Haque}},\ }\href
  {\doibase 10.1103/PhysRevE.89.042112} {\bibfield  {journal} {\bibinfo
  {journal} {Phys. Rev. E}\ }\textbf {\bibinfo {volume} {89}},\ \bibinfo
  {pages} {042112} (\bibinfo {year} {2014})}\BibitemShut {NoStop}%
\bibitem [{\citenamefont {Beugeling}\ \emph {et~al.}(2015)\citenamefont
  {Beugeling}, \citenamefont {Moessner},\ and\ \citenamefont
  {Haque}}]{Beugeling2015}%
  \BibitemOpen
  \bibfield  {author} {\bibinfo {author} {\bibfnamefont {W.}~\bibnamefont
  {Beugeling}}, \bibinfo {author} {\bibfnamefont {R.}~\bibnamefont {Moessner}},
  \ and\ \bibinfo {author} {\bibfnamefont {M.}~\bibnamefont {Haque}},\ }\href
  {\doibase 10.1103/PhysRevE.91.012144} {\bibfield  {journal} {\bibinfo
  {journal} {Phys. Rev. E}\ }\textbf {\bibinfo {volume} {91}},\ \bibinfo
  {pages} {012144} (\bibinfo {year} {2015})}\BibitemShut {NoStop}%
\bibitem [{\citenamefont {Hunter-Jones}\ \emph {et~al.}(2018)\citenamefont
  {Hunter-Jones}, \citenamefont {Liu},\ and\ \citenamefont
  {Zhou}}]{Hunter-Jones2018}%
  \BibitemOpen
  \bibfield  {author} {\bibinfo {author} {\bibfnamefont {N.}~\bibnamefont
  {Hunter-Jones}}, \bibinfo {author} {\bibfnamefont {J.}~\bibnamefont {Liu}}, \
  and\ \bibinfo {author} {\bibfnamefont {Y.}~\bibnamefont {Zhou}},\ }\href
  {https://link.springer.com/article/10.1007/JHEP02(2018)142} {\bibfield
  {journal} {\bibinfo  {journal} {Journal of High Energy Physics}\ }\textbf
  {\bibinfo {volume} {2018}},\ \bibinfo {pages} {142} (\bibinfo {year}
  {2018})}\BibitemShut {NoStop}%
\bibitem [{\citenamefont {Mondaini}\ \emph {et~al.}(2016)\citenamefont
  {Mondaini}, \citenamefont {Fratus}, \citenamefont {Srednicki},\ and\
  \citenamefont {Rigol}}]{Mondaini2016}%
  \BibitemOpen
  \bibfield  {author} {\bibinfo {author} {\bibfnamefont {R.}~\bibnamefont
  {Mondaini}}, \bibinfo {author} {\bibfnamefont {K.~R.}\ \bibnamefont
  {Fratus}}, \bibinfo {author} {\bibfnamefont {M.}~\bibnamefont {Srednicki}}, \
  and\ \bibinfo {author} {\bibfnamefont {M.}~\bibnamefont {Rigol}},\ }\href
  {\doibase 10.1103/PhysRevE.93.032104} {\bibfield  {journal} {\bibinfo
  {journal} {Phys. Rev. E}\ }\textbf {\bibinfo {volume} {93}},\ \bibinfo
  {pages} {032104} (\bibinfo {year} {2016})}\BibitemShut {NoStop}%
\bibitem [{\citenamefont {Yoshizawa}\ \emph {et~al.}(2018)\citenamefont
  {Yoshizawa}, \citenamefont {Iyoda},\ and\ \citenamefont
  {Sagawa}}]{Yoshizawa2018}%
  \BibitemOpen
  \bibfield  {author} {\bibinfo {author} {\bibfnamefont {T.}~\bibnamefont
  {Yoshizawa}}, \bibinfo {author} {\bibfnamefont {E.}~\bibnamefont {Iyoda}}, \
  and\ \bibinfo {author} {\bibfnamefont {T.}~\bibnamefont {Sagawa}},\ }\href
  {\doibase 10.1103/PhysRevLett.120.200604} {\bibfield  {journal} {\bibinfo
  {journal} {Phys. Rev. Lett.}\ }\textbf {\bibinfo {volume} {120}},\ \bibinfo
  {pages} {200604} (\bibinfo {year} {2018})}\BibitemShut {NoStop}%
\bibitem [{\citenamefont {Reimann}(2016)}]{Reimann2016}%
  \BibitemOpen
  \bibfield  {author} {\bibinfo {author} {\bibfnamefont {P.}~\bibnamefont
  {Reimann}},\ }\href {https://www.nature.com/articles/ncomms10821.pdf}
  {\bibfield  {journal} {\bibinfo  {journal} {Nature Communications}\ }\textbf
  {\bibinfo {volume} {7}},\ \bibinfo {pages} {10821} (\bibinfo {year}
  {2016})}\BibitemShut {NoStop}%
\bibitem [{\citenamefont {Borgonovi}\ \emph {et~al.}(2019)\citenamefont
  {Borgonovi}, \citenamefont {Izrailev},\ and\ \citenamefont
  {Santos}}]{Borgonovi2018}%
  \BibitemOpen
  \bibfield  {author} {\bibinfo {author} {\bibfnamefont {F.}~\bibnamefont
  {Borgonovi}}, \bibinfo {author} {\bibfnamefont {F.~M.}\ \bibnamefont
  {Izrailev}}, \ and\ \bibinfo {author} {\bibfnamefont {L.~F.}\ \bibnamefont
  {Santos}},\ }\href {https://link.aps.org/doi/10.1103/PhysRevE.99.010101
  http://arxiv.org/abs/1802.08265} {\bibfield  {journal} {\bibinfo  {journal}
  {Physical Review E}\ }\textbf {\bibinfo {volume} {99}},\ \bibinfo {pages}
  {010101(R)} (\bibinfo {year} {2019})}\BibitemShut {NoStop}%
\bibitem [{\citenamefont {Richter}\ \emph {et~al.}(2018)\citenamefont
  {Richter}, \citenamefont {Gemmer},\ and\ \citenamefont
  {Steinigeweg}}]{Richter2018}%
  \BibitemOpen
  \bibfield  {author} {\bibinfo {author} {\bibfnamefont {J.}~\bibnamefont
  {Richter}}, \bibinfo {author} {\bibfnamefont {J.}~\bibnamefont {Gemmer}}, \
  and\ \bibinfo {author} {\bibfnamefont {R.}~\bibnamefont {Steinigeweg}},\
  }\href {https://arxiv.org/pdf/1805.11625.pdf} {\  (\bibinfo {year} {2018})},\
  \Eprint {http://arxiv.org/abs/1805.11625} {arXiv:1805.11625} \BibitemShut
  {NoStop}%
\bibitem [{\citenamefont {Garc\'{\i}a-Pintos}\ \emph
  {et~al.}(2017)\citenamefont {Garc\'{\i}a-Pintos}, \citenamefont {Linden},
  \citenamefont {Malabarba}, \citenamefont {Short},\ and\ \citenamefont
  {Winter}}]{Garcia-Pintos2015}%
  \BibitemOpen
  \bibfield  {author} {\bibinfo {author} {\bibfnamefont {L.~P.}\ \bibnamefont
  {Garc\'{\i}a-Pintos}}, \bibinfo {author} {\bibfnamefont {N.}~\bibnamefont
  {Linden}}, \bibinfo {author} {\bibfnamefont {A.~S.~L.}\ \bibnamefont
  {Malabarba}}, \bibinfo {author} {\bibfnamefont {A.~J.}\ \bibnamefont
  {Short}}, \ and\ \bibinfo {author} {\bibfnamefont {A.}~\bibnamefont
  {Winter}},\ }\href {\doibase 10.1103/PhysRevX.7.031027} {\bibfield  {journal}
  {\bibinfo  {journal} {Physical Review X}\ }\textbf {\bibinfo {volume} {7}},\
  \bibinfo {pages} {031027} (\bibinfo {year} {2017})}\BibitemShut {NoStop}%
\bibitem [{\citenamefont {Reimann}(2015)}]{Reimann2015}%
  \BibitemOpen
  \bibfield  {author} {\bibinfo {author} {\bibfnamefont {P.}~\bibnamefont
  {Reimann}},\ }\href@noop {} {\bibfield  {journal} {\bibinfo  {journal} {New
  J. Phys.}\ }\textbf {\bibinfo {volume} {17}},\ \bibinfo {pages} {055025}
  (\bibinfo {year} {2015})}\BibitemShut {NoStop}%
\bibitem [{\citenamefont {Nation}\ and\ \citenamefont
  {Porras}(2018)}]{Nation2018}%
  \BibitemOpen
  \bibfield  {author} {\bibinfo {author} {\bibfnamefont {C.}~\bibnamefont
  {Nation}}\ and\ \bibinfo {author} {\bibfnamefont {D.}~\bibnamefont
  {Porras}},\ }\href
  {http://iopscience.iop.org/article/10.1088/1367-2630/aae28f} {\bibfield
  {journal} {\bibinfo  {journal} {New J. Phys.}\ }\textbf {\bibinfo {volume}
  {20}},\ \bibinfo {pages} {103003} (\bibinfo {year} {2018})}\BibitemShut
  {NoStop}%
\bibitem [{\citenamefont {Kubo}(1966)}]{Kubo1966}%
  \BibitemOpen
  \bibfield  {author} {\bibinfo {author} {\bibfnamefont {R.}~\bibnamefont
  {Kubo}},\ }\href {\doibase 10.1088/0034-4885/29/1/306} {\bibfield  {journal}
  {\bibinfo  {journal} {Rep. Prog. Phys.}\ }\textbf {\bibinfo {volume} {29}},\
  \bibinfo {pages} {255} (\bibinfo {year} {1966})}\BibitemShut {NoStop}%
\bibitem [{\citenamefont {Breuer}\ and\ \citenamefont
  {Petruccione}(2002)}]{Breuer2002}%
  \BibitemOpen
  \bibfield  {author} {\bibinfo {author} {\bibfnamefont {H.~P.}\ \bibnamefont
  {Breuer}}\ and\ \bibinfo {author} {\bibfnamefont {F.}~\bibnamefont
  {Petruccione}},\ }\href@noop {} {\emph {\bibinfo {title} {The theory of open
  quantum systems}}}\ (\bibinfo  {publisher} {Oxford University Press},\
  \bibinfo {address} {Great Clarendon Street},\ \bibinfo {year}
  {2002})\BibitemShut {NoStop}%
\bibitem [{\citenamefont {Khatami}\ \emph {et~al.}(2013)\citenamefont
  {Khatami}, \citenamefont {Pupillo}, \citenamefont {Srednicki},\ and\
  \citenamefont {Rigol}}]{Khatami2013}%
  \BibitemOpen
  \bibfield  {author} {\bibinfo {author} {\bibfnamefont {E.}~\bibnamefont
  {Khatami}}, \bibinfo {author} {\bibfnamefont {G.}~\bibnamefont {Pupillo}},
  \bibinfo {author} {\bibfnamefont {M.}~\bibnamefont {Srednicki}}, \ and\
  \bibinfo {author} {\bibfnamefont {M.}~\bibnamefont {Rigol}},\ }\href
  {\doibase 10.1103/PhysRevLett.111.050403} {\bibfield  {journal} {\bibinfo
  {journal} {Phys. Rev. Lett.}\ }\textbf {\bibinfo {volume} {111}},\ \bibinfo
  {pages} {050403} (\bibinfo {year} {2013})}\BibitemShut {NoStop}%
\bibitem [{\citenamefont {Deutsch}(1991{\natexlab{b}})}]{Deutscha}%
  \BibitemOpen
  \bibfield  {author} {\bibinfo {author} {\bibfnamefont {J.~M.}\ \bibnamefont
  {Deutsch}},\ }\href@noop {} {\bibfield  {journal} {\bibinfo  {journal}
  {(unpublished)}\ } (\bibinfo {year} {1991}{\natexlab{b}})}\BibitemShut
  {NoStop}%
\bibitem [{Foo()}]{Footnote}%
  \BibitemOpen
  \href@noop {} {\bibinfo  {journal} {We note that this differs from Deutsch's
  initial calculation by a factor of two, this is analytically and numerically
  shown in [31]}\ }\BibitemShut {NoStop}%
\bibitem [{\citenamefont {Guhr}\ \emph {et~al.}(1998)\citenamefont {Guhr},
  \citenamefont {Müller–Groeling},\ and\ \citenamefont
  {Weidenmüller}}]{Guhr1998}%
  \BibitemOpen
\bibfield  {journal} {  }\bibfield  {author} {\bibinfo {author} {\bibfnamefont
  {T.}~\bibnamefont {Guhr}}, \bibinfo {author} {\bibfnamefont {A.}~\bibnamefont
  {Müller–Groeling}}, \ and\ \bibinfo {author} {\bibfnamefont {H.~A.}\
  \bibnamefont {Weidenmüller}},\ }\href {\doibase
  https://doi.org/10.1016/S0370-1573(97)00088-4} {\bibfield  {journal}
  {\bibinfo  {journal} {Physics Reports}\ }\textbf {\bibinfo {volume} {299}},\
  \bibinfo {pages} {189 } (\bibinfo {year} {1998})}\BibitemShut {NoStop}%
\bibitem [{\citenamefont {Mondaini}\ and\ \citenamefont
  {Rigol}(2017)}]{Mondaini2017}%
  \BibitemOpen
  \bibfield  {author} {\bibinfo {author} {\bibfnamefont {R.}~\bibnamefont
  {Mondaini}}\ and\ \bibinfo {author} {\bibfnamefont {M.}~\bibnamefont
  {Rigol}},\ }\href {\doibase 10.1103/PhysRevE.96.012157} {\bibfield  {journal}
  {\bibinfo  {journal} {Phys. Rev. E}\ }\textbf {\bibinfo {volume} {96}},\
  \bibinfo {pages} {012157} (\bibinfo {year} {2017})}\BibitemShut {NoStop}%
\bibitem [{\citenamefont {Santos}\ \emph {et~al.}(2012)\citenamefont {Santos},
  \citenamefont {Borgonovi},\ and\ \citenamefont {Izrailev}}]{Santos2012}%
  \BibitemOpen
  \bibfield  {author} {\bibinfo {author} {\bibfnamefont {L.~F.}\ \bibnamefont
  {Santos}}, \bibinfo {author} {\bibfnamefont {F.}~\bibnamefont {Borgonovi}}, \
  and\ \bibinfo {author} {\bibfnamefont {F.~M.}\ \bibnamefont {Izrailev}},\
  }\href {\doibase 10.1103/PhysRevE.85.036209} {\bibfield  {journal} {\bibinfo
  {journal} {Phys. Rev. E}\ }\textbf {\bibinfo {volume} {85}},\ \bibinfo
  {pages} {036209} (\bibinfo {year} {2012})}\BibitemShut {NoStop}%
\bibitem [{\citenamefont {Atas}\ and\ \citenamefont
  {Bogomolny}(2017)}]{Atlas2017}%
  \BibitemOpen
  \bibfield  {author} {\bibinfo {author} {\bibfnamefont {Y.~Y.}\ \bibnamefont
  {Atas}}\ and\ \bibinfo {author} {\bibfnamefont {E.}~\bibnamefont
  {Bogomolny}},\ }\href {\doibase 10.1088/1751-8121/aa81f6} {\bibfield
  {journal} {\bibinfo  {journal} {J. Phys. A: Mathematical and Theoretical}\
  }\textbf {\bibinfo {volume} {50}},\ \bibinfo {pages} {385102} (\bibinfo
  {year} {2017})}\BibitemShut {NoStop}%
\bibitem [{\citenamefont {Borgonovi}\ \emph {et~al.}(2017)\citenamefont
  {Borgonovi}, \citenamefont {Mattiotti},\ and\ \citenamefont
  {Izrailev}}]{Borgonovi2017}%
  \BibitemOpen
  \bibfield  {author} {\bibinfo {author} {\bibfnamefont {F.}~\bibnamefont
  {Borgonovi}}, \bibinfo {author} {\bibfnamefont {F.}~\bibnamefont
  {Mattiotti}}, \ and\ \bibinfo {author} {\bibfnamefont {F.~M.}\ \bibnamefont
  {Izrailev}},\ }\href
  {https://journals.aps.org/pre/pdf/10.1103/PhysRevE.95.042135} {\bibfield
  {journal} {\bibinfo  {journal} {Physical Review E}\ }\textbf {\bibinfo
  {volume} {95}},\ \bibinfo {pages} {042135} (\bibinfo {year}
  {2017})}\BibitemShut {NoStop}%
\bibitem [{\citenamefont {Anza}\ \emph {et~al.}(2018)\citenamefont {Anza},
  \citenamefont {Gogolin},\ and\ \citenamefont {Huber}}]{Anza2018}%
  \BibitemOpen
  \bibfield  {author} {\bibinfo {author} {\bibfnamefont {F.}~\bibnamefont
  {Anza}}, \bibinfo {author} {\bibfnamefont {C.}~\bibnamefont {Gogolin}}, \
  and\ \bibinfo {author} {\bibfnamefont {M.}~\bibnamefont {Huber}},\ }\href
  {https://journals.aps.org/prl/pdf/10.1103/PhysRevLett.120.150603} {\bibfield
  {journal} {\bibinfo  {journal} {Physical Review Letters}\ }\textbf {\bibinfo
  {volume} {120}},\ \bibinfo {pages} {150603} (\bibinfo {year}
  {2018})}\BibitemShut {NoStop}%
\bibitem [{\citenamefont {Merali}(2017)}]{Merali2017}%
  \BibitemOpen
  \bibfield  {author} {\bibinfo {author} {\bibfnamefont {Z.}~\bibnamefont
  {Merali}},\ }\href {\doibase 10.1038/551020a} {\bibfield  {journal} {\bibinfo
   {journal} {Nature}\ }\textbf {\bibinfo {volume} {551}},\ \bibinfo {pages}
  {20} (\bibinfo {year} {2017})}\BibitemShut {NoStop}%
\end{thebibliography}%
\end{document}